%% file: main.tex
\documentclass[12pt,a4paper]{article}

\usepackage{ifthen} 
\usepackage{float}
\usepackage{textcomp}
\usepackage{rotating}
\usepackage{lscape}
\newboolean{pdflatex}
\setboolean{pdflatex}{true} 

\newboolean{articletitles}
\setboolean{articletitles}{true} 

\newboolean{uprightparticles}
\setboolean{uprightparticles}{false} 

\newboolean{inbibliography}
\setboolean{inbibliography}{false} 

\input{preamble}
\input{mydefine}

\usepackage{longtable} 

\begin{document}

\renewcommand{\thefootnote}{\fnsymbol{footnote}}
\setcounter{footnote}{1}

\input{title-LHCb-PAPER}

\renewcommand{\thefootnote}{\arabic{footnote}}
\setcounter{footnote}{0}

\pagestyle{plain} 
\setcounter{page}{1}
\pagenumbering{arabic}

\input{introduction}

\input{detector}

\input{selection}

\input{X_determination}

\input{A_determination}

\input{systematics}

\input{result}
\input{acknowledgements}
\input{app_result}

\addcontentsline{toc}{section}{References}
\setboolean{inbibliography}{true}
\bibliographystyle{LHCb}
\bibliography{main,LHCb-PAPER,LHCb-CONF,LHCb-DP,LHCb-TDR,MyBib}

\newpage

\input{LHCb_HD_authorlist_2015-06-23}

\newpage

\end{document}

%% file: preamble.tex
\usepackage{microtype}
\usepackage{lineno}  
\usepackage{xspace} 
\usepackage{caption} 

\usepackage{graphicx}  
\usepackage{color}
\usepackage{colortbl}
\graphicspath{{./figs/}} 

\usepackage{amsmath} 
\usepackage{amssymb}
\usepackage{amsfonts}
\usepackage{upgreek} 

\newcommand*\patchAmsMathEnvironmentForLineno[1]{%
\expandafter\let\csname old#1\expandafter\endcsname\csname #1\endcsname
\expandafter\let\csname oldend#1\expandafter\endcsname\csname
end#1\endcsname
 \renewenvironment{#1}%
   {\linenomath\csname old#1\endcsname}%
   {\csname oldend#1\endcsname\endlinenomath}%
}
\newcommand*\patchBothAmsMathEnvironmentsForLineno[1]{%
  \patchAmsMathEnvironmentForLineno{#1}%
  \patchAmsMathEnvironmentForLineno{#1*}%
}
\AtBeginDocument{%
\patchBothAmsMathEnvironmentsForLineno{equation}%
\patchBothAmsMathEnvironmentsForLineno{align}%
\patchBothAmsMathEnvironmentsForLineno{flalign}%
\patchBothAmsMathEnvironmentsForLineno{alignat}%
\patchBothAmsMathEnvironmentsForLineno{gather}%
\patchBothAmsMathEnvironmentsForLineno{multline}%
\patchBothAmsMathEnvironmentsForLineno{eqnarray}%
}

\usepackage{hyperref}    
\usepackage[all]{hypcap} 

\input{lhcb-symbols-def} 

\usepackage{cite} 
\usepackage{mciteplus}

%% file: lhcb-symbols-def.tex
\def\lhcb {\mbox{LHCb}\xspace}

\def\cms    {\mbox{CMS}\xspace}

\def\belle  {\mbox{Belle}\xspace}

\def\MagUp {\mbox{\em Mag\kern -0.05em Up}\xspace}

\ifthenelse{\boolean{uprightparticles}}%
{

 \def\Pmu         {\ensuremath{\upmu}\xspace}

 \def\Ppi         {\ensuremath{\uppi}\xspace}

 \def\Ppsi        {\ensuremath{\uppsi}\xspace}

 \def\PDelta      {\ensuremath{\Delta}\xspace}                 
 \def\PXi      {\ensuremath{\Xi}\xspace}                 
 \def\PLambda      {\ensuremath{\Lambda}\xspace}                 
 \def\PSigma      {\ensuremath{\Sigma}\xspace}                 
 \def\POmega      {\ensuremath{\Omega}\xspace}                 
 \def\PUpsilon      {\ensuremath{\Upsilon}\xspace}                 
 
 \def\PB      {\ensuremath{\mathrm{B}}\xspace}                 
                  
 \def\PD      {\ensuremath{\mathrm{D}}\xspace}

 \def\PJ      {\ensuremath{\mathrm{J}}\xspace}                 
 \def\PK      {\ensuremath{\mathrm{K}}\xspace}

 \def\Pb      {\ensuremath{\mathrm{b}}\xspace}                 
 \def\Pc      {\ensuremath{\mathrm{c}}\xspace}

 \def\Pi      {\ensuremath{\mathrm{i}}\xspace}

 \def\Pp      {\ensuremath{\mathrm{p}}\xspace}

 \def\Ps      {\ensuremath{\mathrm{s}}\xspace}

}
{

 \def\Pmu         {\ensuremath{\mu}\xspace}

 \def\Ppi         {\ensuremath{\pi}\xspace}

 \def\Ppsi        {\ensuremath{\psi}\xspace}                 
                  
 \mathchardef\PDelta="7101
 \mathchardef\PXi="7104
 \mathchardef\PLambda="7103
 \mathchardef\PSigma="7106
 \mathchardef\POmega="710A
 \mathchardef\PUpsilon="7107
                  
 \def\PB      {\ensuremath{B}\xspace}                 
                  
 \def\PD      {\ensuremath{D}\xspace}

 \def\PJ      {\ensuremath{J}\xspace}                 
 \def\PK      {\ensuremath{K}\xspace}

 \def\Pb      {\ensuremath{b}\xspace}                 
 \def\Pc      {\ensuremath{c}\xspace}

 \def\Pi      {\ensuremath{i}\xspace}

 \def\Pp      {\ensuremath{p}\xspace}

 \def\Ps      {\ensuremath{s}\xspace}

}

\makeatletter
\ifcase \@ptsize \relax
  \newcommand{\miniscule}{\@setfontsize\miniscule{4}{5}}
\or
  \newcommand{\miniscule}{\@setfontsize\miniscule{5}{6}}
\or
  \newcommand{\miniscule}{\@setfontsize\miniscule{5}{6}}
\fi
\makeatother

\DeclareRobustCommand{\optbar}[1]{\shortstack{{\miniscule (\rule[.5ex]{1.25em}{.18mm})}
  \\ [-.7ex] $#1$}}

\def\mumu       {{\ensuremath{\Pmu^+\Pmu^-}}\xspace}

\def\squark    {{\ensuremath{\Ps}}\xspace}

\def\cquark    {{\ensuremath{\Pc}}\xspace}

\def\bquark    {{\ensuremath{\Pb}}\xspace}

\def\pion   {{\ensuremath{\Ppi}}\xspace}

\def\pip    {{\ensuremath{\pion^+}}\xspace}
\def\pim    {{\ensuremath{\pion^-}}\xspace}

\def\kaon    {{\ensuremath{\PK}}\xspace}
  \def\Kbar    {{\kern 0.2em\overline{\kern -0.2em \PK}{}}\xspace}

\def\KorKbar    {\kern 0.18em\optbar{\kern -0.18em K}{}\xspace}

\def\Kp      {{\ensuremath{\kaon^+}}\xspace}
\def\Km      {{\ensuremath{\kaon^-}}\xspace}

\def\Kstarzb {{\ensuremath{\Kbar{}^{*0}}}\xspace}

\def\Kstarb  {{\ensuremath{\Kbar{}^*}}\xspace}

  \def\Dbar    {{\kern 0.2em\overline{\kern -0.2em \PD}{}}\xspace}
\def\D       {{\ensuremath{\PD}}\xspace}

\def\DorDbar    {\kern 0.18em\optbar{\kern -0.18em D}{}\xspace}
\def\Dz      {{\ensuremath{\D^0}}\xspace}

\def\Dp      {{\ensuremath{\D^+}}\xspace}

\def\Dstarp  {{\ensuremath{\D^{*+}}}\xspace}

\def\B       {{\ensuremath{\PB}}\xspace}
\def\Bbar    {{\ensuremath{\kern 0.18em\overline{\kern -0.18em \PB}{}}}\xspace}

\def\BorBbar    {\kern 0.18em\optbar{\kern -0.18em B}{}\xspace}

\def\Bs      {{\ensuremath{\B^0_\squark}}\xspace}
\def\Bsb     {{\ensuremath{\Bbar{}^0_\squark}}\xspace}
\def\Bdb     {{\ensuremath{\Bbar{}^0}}\xspace}

\def\jpsi     {{\ensuremath{{\PJ\mskip -3mu/\mskip -2mu\Ppsi\mskip 2mu}}}\xspace}

  \def\Y#1S{\ensuremath{\PUpsilon{(#1S)}}\xspace}

\def\proton      {{\ensuremath{\Pp}}\xspace}
\def\antiproton  {{\ensuremath{\overline \proton}}\xspace}

\def\Lz          {{\ensuremath{\PLambda}}\xspace}
\def\Lbar        {{\ensuremath{\kern 0.1em\overline{\kern -0.1em\PLambda}}}\xspace}
\def\LorLbar    {\kern 0.18em\optbar{\kern -0.18em \PLambda}{}\xspace}

\def\Lb      {{\ensuremath{\Lz^0_\bquark}}\xspace}
\def\Lbbar   {{\ensuremath{\Lbar{}^0_\bquark}}\xspace}
\def\Lc      {{\ensuremath{\Lz^+_\cquark}}\xspace}

\def\BF         {{\ensuremath{\cal B}}\xspace}

\def\BR         {\BF}

\def\to                 {\ensuremath{\rightarrow}\xspace}

\def\eps   {{\ensuremath{\varepsilon}}\xspace}

\def\AT#1     {\ensuremath{A_{\mathrm{T}}^{#1}}\xspace}           

\def\C#1      {\ensuremath{\mathcal{C}_{#1}}\xspace}                       
\def\Cp#1     {\ensuremath{\mathcal{C}_{#1}^{'}}\xspace}                    
\def\Ceff#1   {\ensuremath{\mathcal{C}_{#1}^{\mathrm{(eff)}}}\xspace}        
\def\Cpeff#1  {\ensuremath{\mathcal{C}_{#1}^{'\mathrm{(eff)}}}\xspace}       
\def\Ope#1    {\ensuremath{\mathcal{O}_{#1}}\xspace}                       
\def\Opep#1   {\ensuremath{\mathcal{O}_{#1}^{'}}\xspace}                    

\newcommand{\tev}{\ifthenelse{\boolean{inbibliography}}{\ensuremath{~T\kern -0.05em eV}\xspace}{\ensuremath{\mathrm{\,Te\kern -0.1em V}}}\xspace}
\newcommand{\gev}{\ensuremath{\mathrm{\,Ge\kern -0.1em V}}\xspace}
\newcommand{\mev}{\ensuremath{\mathrm{\,Me\kern -0.1em V}}\xspace}
\newcommand{\kev}{\ensuremath{\mathrm{\,ke\kern -0.1em V}}\xspace}
\newcommand{\ev}{\ensuremath{\mathrm{\,e\kern -0.1em V}}\xspace}
\newcommand{\gevc}{\ensuremath{{\mathrm{\,Ge\kern -0.1em V\!/}c}}\xspace}
\newcommand{\mevc}{\ensuremath{{\mathrm{\,Me\kern -0.1em V\!/}c}}\xspace}
\newcommand{\gevcc}{\ensuremath{{\mathrm{\,Ge\kern -0.1em V\!/}c^2}}\xspace}
\newcommand{\gevgevcccc}{\ensuremath{{\mathrm{\,Ge\kern -0.1em V^2\!/}c^4}}\xspace}
\newcommand{\mevcc}{\ensuremath{{\mathrm{\,Me\kern -0.1em V\!/}c^2}}\xspace}

\def\mum  {\ensuremath{{\,\upmu\rm m}}\xspace}

\def\nb {\ensuremath{\rm \,nb}\xspace}

\def\invpb {\ensuremath{\mbox{\,pb}^{-1}}\xspace}

\def\invfb   {\ensuremath{\mbox{\,fb}^{-1}}\xspace}

\def\ps   {\ensuremath{{\rm \,ps}}\xspace}

\newcommand{\stat}{\ensuremath{\mathrm{\,(stat)}}\xspace}
\newcommand{\syst}{\ensuremath{\mathrm{\,(syst)}}\xspace}

\newcommand{\chisq}{\ensuremath{\chi^2}\xspace}
\newcommand{\chisqndf}{\ensuremath{\chi^2/\mathrm{ndf}}\xspace}

\def\deriv {\ensuremath{\mathrm{d}}}

\def\gsim{{~\raise.15em\hbox{$>$}\kern-.85em
          \lower.35em\hbox{$\sim$}~}\xspace}
\def\lsim{{~\raise.15em\hbox{$<$}\kern-.85em
          \lower.35em\hbox{$\sim$}~}\xspace}

\newcommand{\mean}[1]{\ensuremath{\left\langle #1 \right\rangle}} 

\def\sPlot{\mbox{\em sPlot}\xspace}

\def\sqs   {\ensuremath{\protect\sqrt{s}}\xspace}

\def\ptot       {\mbox{$p$}\xspace}
\def\pt         {\mbox{$p_{\rm T}$}\xspace}

\newcommand{\lum} {\ensuremath{\mathcal{L}}\xspace}

\def\evtgen     {\mbox{\textsc{EvtGen}}\xspace}

\def\geant      {\mbox{\textsc{Geant4}}\xspace}

\def\photos     {\mbox{\textsc{Photos}}\xspace}

\def\pythia     {\mbox{\textsc{Pythia}}\xspace}

\def\tell1  {TELL1\xspace}
\def\ukl1   {UKL1\xspace}

\newcommand{\eg}{\mbox{\itshape e.g.}\xspace}

%% file: mydefine.tex
\def\mypb {\ensuremath{\rm pb}\xspace}

\newcommand{\mygevc}{\ensuremath{{\mathrm{Ge\kern -0.1em V\!/}c}}\xspace}
\newcommand{\xx}{\ensuremath{\kern 0.5em }}
\def\y {\ensuremath{y}\xspace}
\def\dy {\ensuremath{\deriv\y}\xspace}
\def\Dy {\ensuremath{\Delta\y}\xspace}
\def\dpt {\ensuremath{\deriv\pt}\xspace}
\def\Dpt {\ensuremath{\Delta\pt}\xspace}
\def\inter{\ensuremath{{\rm inter}}\xspace}

\def\ndf{\ensuremath{{\rm ndf}}\xspace}
\def\NsigLb {\ensuremath{N_{\rm sig}^{\Lb}}\xspace}
\def\NsigBdb {\ensuremath{N_{\rm sig}^{\Bdb}}\xspace}
\def\epsLb {\ensuremath{\eps_{\rm tot}^{\Lb}}\xspace}
\def\epsBdb {\ensuremath{\eps_{\rm tot}^{\Bdb}}\xspace}
\def\RLbBdb{\ensuremath{R_{\Lb/\Bdb}}\xspace}
\def\fLbd{\ensuremath{f_{\Lb}/f_d}\xspace}
\def\fLbud{\ensuremath{f_{\Lb}/(f_u+f_d)}\xspace}
\def\apd{\ensuremath{a_{\rm p+d}}\xspace}
\def\aprod{\ensuremath{a_{\rm prod}}\xspace}
\def\adecay{\ensuremath{a_{\rm decay}}\xspace}
\def\aDproton{\ensuremath{a_{\rm D}^{p}}\xspace}
\def\aDKaon{\ensuremath{a_{\rm D}^{K}}\xspace}
\def\aPID{\ensuremath{a_{\rm PID}}\xspace}
\def\Araw{\ensuremath{A_{\rm raw}}\xspace}

\newcommand{\Lbpk}{\ensuremath{\Lb\to\jpsi\proton\Km}\xspace}
\newcommand{\antiLbpk}{\ensuremath{\Lbbar\to\jpsi\antiproton\Kp}\xspace}
\newcommand{\Bpik}{\ensuremath{\Bdb\to\jpsi\Kstarzb}\xspace}
\newcommand{\BJpsiKpi}{\ensuremath{\Bdb\to\jpsi\Kstarb(892)^0}\xspace}

\newcommand{\LbLcpi}{\ensuremath{\Lb\to\Lc\pim}\xspace}
\newcommand{\LbJpsiLambda}{\ensuremath{\Lb\to\jpsi\Lz}\xspace}
\newcommand{\LbJpsippi}{\ensuremath{\Lb\to\jpsi\proton\pim}\xspace}

\newcommand{\Jpsimumu}{\ensuremath{\jpsi\to\mumu}\xspace}
\newcommand{\KstarzbKpi}{\ensuremath{\Kstarzb\to\Km\pip}\xspace}
\newcommand{\BdbDpi}{\ensuremath{\Bdb\to\Dp\pim}\xspace}

\def\pp {\ensuremath{pp}\xspace}

\def\sPlot{\mbox{\em sPlot}\xspace}

\usepackage{multirow} 
\usepackage{booktabs} 
\usepackage{rotating}

%% file: title-LHCb-PAPER.tex
\begin{titlepage}
\pagenumbering{roman}

\vspace*{-1.5cm}
\centerline{\large EUROPEAN ORGANIZATION FOR NUCLEAR RESEARCH (CERN)}
\vspace*{0.9cm}
\noindent
\begin{tabular*}{\linewidth}{lc@{\extracolsep{\fill}}r@{\extracolsep{0pt}}}
\ifthenelse{\boolean{pdflatex}}
{\vspace*{-2.7cm}\mbox{\!\!\!\includegraphics[width=.14\textwidth]{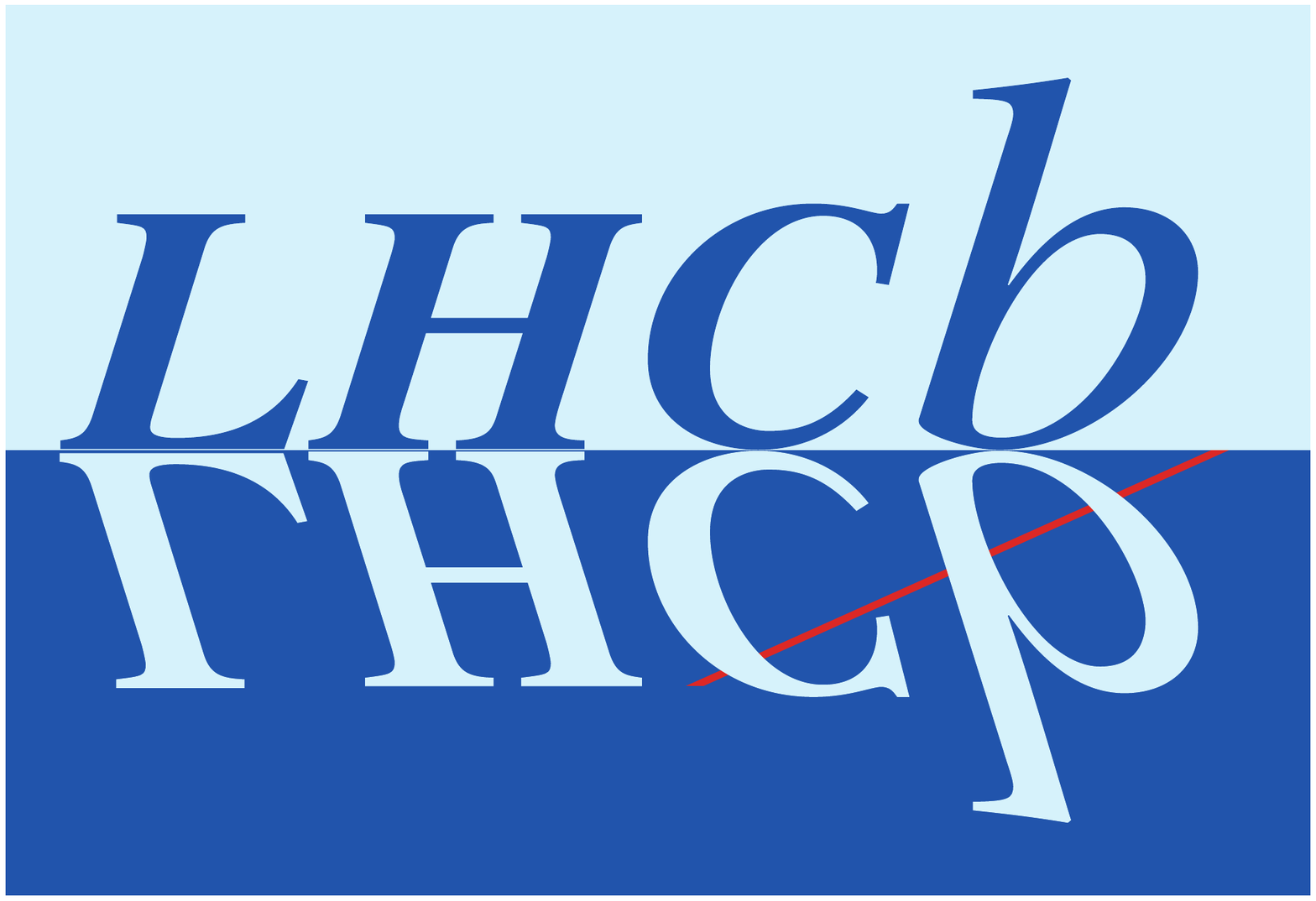}} & &}%
{\vspace*{-1.2cm}\mbox{\!\!\!\includegraphics[width=.12\textwidth]{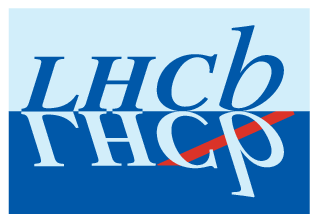}} & &}%
\\
 & & CERN-PH-EP-2015-223 \\   
 & & LHCb-PAPER-2015-032 \\   
 & & 25 January 2016 \\ 
\end{tabular*}

\vspace*{1.1cm}

{\bf\boldmath\Huge
\begin{center}
  Study of the production of $\Lb$ and $\Bdb$ hadrons in $pp$ collisions
  and first measurement of the $\Lbpk$ branching fraction
\end{center}
}

\vspace*{0.1cm}

\begin{center}
The LHCb collaboration\footnote{Authors are listed at the end of this paper.}
\end{center}

\vspace{\fill}

\begin{abstract}
  \noindent
The product of the $\Lb$ ($\Bdb$) differential production cross-section 
and the branching fraction of the decay $\Lbpk$ ($\BJpsiKpi$) is measured
as a function of the beauty hadron transverse momentum, $\pt$, and rapidity, $y$.
The kinematic region of the measurements is $\pt<20\gevc$ and $2.0<y<4.5$.
The measurements use a data sample corresponding to an integrated luminosity of $3\invfb$ collected by the \lhcb detector in $pp$ collisions at centre-of-mass energies $\sqrt{s}=7\tev$ in 2011 and $\sqrt{s}=8\tev$ in 2012.
Based on previous LHCb results of the fragmentation fraction ratio,
$\fLbd$, the branching fraction of the decay $\Lbpk$ is measured to be 
\begin{equation*}
\BR(\Lbpk)=
\input{table/number/branching_fraction/res.tex},
\end{equation*}
where the first uncertainty is statistical, the second is systematic, 
the third is due to the uncertainty on the branching fraction of the decay $\BJpsiKpi$, 
and the fourth is due to the knowledge of \fLbd.
The sum of the asymmetries in the production and decay between $\Lb$ and $\Lbbar$ is 
also measured as a function of $\pt$ and $y$.
The previously published branching fraction of \LbJpsippi, relative to that of \Lbpk, is updated.
The branching fractions of $\Lb\to{}P_c^+(\to\jpsi\proton)\Km$ are determined.
  
\end{abstract}

\vspace*{0.5cm}

\begin{center}
  Published in Chin. Phys. C40 (2016) 011001
\end{center}

\vspace*{0.5cm}

{\footnotesize 
\centerline{\copyright~CERN on behalf of the \lhcb collaboration, licence \href{http://creativecommons.org/licenses/by/4.0/}{CC-BY-4.0}.}}
\vspace*{2mm}

\end{titlepage}

\newpage
\setcounter{page}{2}
\mbox{~}

\cleardoublepage

%% file: table/number/branching_fraction/res.tex
(3.17 \pm 0.04 \pm 0.07 \pm 0.34^{+0.45}_{-0.28}) \times 10^{-4}

%% file: introduction.tex
\section{Introduction}
\label{sec:Introduction}

In quantum chromodynamics (QCD) the production process of $b$ hadrons can be divided into two steps, assuming factorisation: a hard process for \bquark production and a soft process to describe hadronisation.
The hard process can be predicted by perturbative calculations in QCD; 
the soft process is parameterised by the fragmentation function, which has large uncertainties due to non-perturbative QCD contributions. 
The study of the production of $\bquark$ hadrons tests the factorisation ansatz.
The ground state of the $b$-baryon family, $\Lb$, has a wide range of decay modes.
The study of its production and decays can offer complementary information 
to that obtained from the study of $B$ mesons.
The kinematic dependence of the production of $\Lb$ baryons relative to that of 
$B$ mesons can test differences in the $\bquark$ quark hadronisation process
between the two~\cite{Lee:2007wr,Oh:2009zj}.
Furthermore, the asymmetry of heavy flavoured baryons and antibaryons 
produced in $pp$ collisions is an important input for various asymmetry measurements.
Leading-order QCD calculations predict equal production cross-sections 
for heavy baryons and heavy anti-baryons, 
while measurements at the ISR showed that \Lc production is favoured in \pp collisions
at forward rapidity, $y$~\cite{Lockman:1979aj,Chauvat:1987kb}.
The \cms experiment measured the \Lb and \Lbbar production ratio in $pp$ collisions 
at $7\tev$, and no asymmetry was observed, but the large uncertainties preclude 
definitive conclusions~\cite{Chatrchyan:2012xg}.
Measurements at \lhcb can provide further tests of existing mechanisms, \eg, 
the string drag effect or the leading quark effect~\cite{Rosner:2014gta}.

Measurements of $\Lb$ production to date have mostly been based on 
semileptonic decays and the hadronic decays $\LbJpsiLambda$ and $\LbLcpi$
(charge-conjugation is implied throughout the paper unless otherwise specified).
Using semileptonic decays, the \lhcb experiment measured the ratio of \Lb 
baryon production to light $B$ meson production, $\fLbud$~\cite{LHCb-PAPER-2011-018}.
The kinematic dependence of the ratio of \Lb to \Bdb production, $\fLbd$, 
was measured using \LbLcpi and \BdbDpi decays, and the absolute branching fraction 
$\BR(\LbLcpi)$ was determined~\cite{LHCb-PAPER-2014-004}. 

In this paper, the \Lb candidates are reconstructed in the decay channel $\Lbpk$, which was first observed by \lhcb in 2013~\cite{LHCb-PAPER-2013-032}.
Compared with the open-charm decays of $\Lb$ baryons, 
this channel has higher trigger efficiencies, 
especially in the region of low transverse momentum, $\pt$.
Two pentaquark-charmonium states $P_c(4380)^+$ and $P_c(4450)^+$ were observed  
by \lhcb~\cite{LHCb-PAPER-2015-029} in the amplitude analysis of the $\Lbpk$ decay. 
The measurement of the absolute branching fraction of \Lbpk in the current paper 
allows the pentaquark branching fractions to be determined.
Other \Lb decays with a charmonium meson in the final state, such as the Cabibbo-suppressed decay \LbJpsippi~\cite{LHCb-PAPER-2014-020}, can use the \Lbpk decay
as a reference to measure their absolute branching fractions.

The product of the $\Lb$ (\Bdb) differential production cross-section
and the branching fraction of the \Lbpk (\Bpik) decay 
is measured as a function of $\pt$ and \y,
where $\Kstarzb$ indicates the $\Kstarb(892)^0$ meson throughout the text.
The kinematic region of these measurements is $\pt<20\gevc$ and $2.0<y<4.5$ for the $b$ hadron.
The production ratio of the two $b$ hadrons, defined as
\begin{equation}
\label{eq1}
\RLbBdb\equiv\frac{\sigma(\Lb)\BR(\Lbpk)}{\sigma(\Bdb)\BR(\Bpik)},
\end{equation}
is determined,
taking advantage of the cancellation of some uncertainties in
both experimental measurements and theoretical calculations.
Here, $\sigma(\Lb)$ and $\sigma(\Bdb)$ represent the production cross-sections of 
$\Lb$ and $\Bdb$ hadrons in $pp$ collisions.
The branching fraction $\BR(\Lbpk)$ is calculated from this result 
using previous measurements of $\fLbd$~\cite{LHCb-PAPER-2011-018,LHCb-PAPER-2014-004}
and $\BF(\Bpik)$~\cite{Abe:2002haa}.
The kinematic dependence of the sum of the asymmetries in the production and decay,
$\apd\equiv\aprod+\adecay$,
between $\Lb$ and $\Lbbar$ is studied using $\Lbpk$ and $\antiLbpk$ decays.
Furthermore, using the measurement of $\BR(\Lbpk)$, the branching fractions of the decays 
$\LbJpsippi$ and $\Lb\to{}P_c^+(\to\jpsi\proton)\Km$ are determined.

The measurements in this paper are based on a data sample corresponding to 
an integrated luminosity of $3\invfb$, 
collected by the \lhcb experiment in $pp$ collisions 
at centre-of-mass energies $\sqs=7\tev$ in 2011 and $\sqs=8\tev$ in 2012.
Separate measurements are performed for each of the two centre-of-mass energies.

%% file: detector.tex
\section{Detector and simulation}
\label{sec:Detector}
The \lhcb detector~\cite{Alves:2008zz,LHCb-DP-2014-002} is a single-arm forward
spectrometer covering the \mbox{pseudorapidity} range $2<\eta <5$,
designed for the study of particles containing \bquark or \cquark
quarks. The detector includes a high-precision tracking system
consisting of a silicon-strip vertex detector surrounding the $pp$
interaction region~\cite{LHCb-DP-2014-001}, a large-area silicon-strip detector located
upstream of a dipole magnet with a bending power of about
$4{\rm\,Tm}$, and three stations of silicon-strip detectors and straw
drift tubes~\cite{LHCb-DP-2013-003} placed downstream of the magnet.
The tracking system provides a measurement of momentum, \ptot, of charged particles with
a relative uncertainty that varies from 0.5\% at low momentum to 1.0\% at 200\gevc.
The minimum distance of a track to a primary vertex (PV), the impact parameter (IP), is measured with a resolution of $(15+29/\pt)\mum$,
where \pt is the component of the momentum transverse to the beam, in\,\gevc.
Different types of charged hadrons are distinguished using information
from two ring-imaging Cherenkov detectors~\cite{LHCb-DP-2012-003}. 
Photons, electrons and hadrons are identified by a calorimeter system consisting of
scintillating-pad and preshower detectors, an electromagnetic
calorimeter and a hadronic calorimeter. Muons are identified by a
system composed of alternating layers of iron and multiwire
proportional chambers~\cite{LHCb-DP-2012-002}.
The online event selection is performed by a trigger~\cite{LHCb-DP-2012-004}, 
which consists of a hardware stage, based on information from the calorimeter and muon
systems, followed by a software stage, which applies a full event
reconstruction.
In the hardware trigger, events are selected by requiring at least one high-$\pt$ track
that is consistent with a muon hypothesis.
In the software trigger, two well-reconstructed muons are required 
to form a vertex with good fit $\chisq$ and to have an invariant mass consistent with
that of the \jpsi meson~\cite{PDG2014}.
The trigger also requires a significant displacement
between the $\jpsi$ vertex and the associated PV of the $pp$ collision.

In the simulation, $pp$ collisions are generated using
\pythia~\cite{Sjostrand:2006za,Sjostrand:2007gs}
 with a specific \lhcb configuration~\cite{LHCb-PROC-2010-056}.  
Decays of hadronic particles are described by \evtgen~\cite{Lange:2001uf}, 
in which final-state radiation is generated using \photos~\cite{Golonka:2005pn}. 
The interaction of the generated particles with the detector, and its response,
are implemented using the \geant toolkit~\cite{Allison:2006ve, Agostinelli:2002hh} 
as described in Ref.~\cite{LHCb-PROC-2011-006}.
The physics models used by \lhcb in \geant for hadronic interactions have been tested against
experimental data from COMPAS \cite{PDG2014}, and good agreement was observed.\footnote{Data files are courtesy of the COMPAS Group, IHEP, Protvino, Russia.}

%% file: selection.tex
\section{Event selection}
\label{sec:selection}

Candidates for $\Lb$ ($\Bdb$) hadrons are reconstructed in the $\Lbpk$ ($\Bpik$) 
decay channel, where the $\jpsi$ mesons are reconstructed in the dimuon final state, 
and $\Kstarzb$ candidates are reconstructed from $\Kstarzb\to\Km\pip$ decays.
Since the $\Lbpk$ and $\Bpik$ decays have the same topology, 
a similar event selection is adopted for both.

An offline selection is applied after the trigger and is divided into two steps: 
a preselection and a multivariate selection based on 
a boosted decision tree (BDT)~\cite{Breiman,Roe,AdaBoost,Hocker:2007ht}.
In the preselection, each track of the \Lb (\Bdb) candidate is required to be of good quality \cite{LHCb-DP-2014-002,Fruhwirth:1987fm,Needham:2007oba,Needham:2007zzb}.
Identified muons are required to have \pt greater than $550\mevc$,
while hadrons are required to have \pt greater than $250\mevc$.
The muons should be inconsistent with originating from any PV, as determined by their impact parameter.
Each $\jpsi$ candidate is required to have a good vertex fit $\chisq$ 
and an invariant mass within ${}_{-48}^{+43}\mevcc$ of the known \jpsi mass~\cite{PDG2014}.
Particle identification (PID) requirements are imposed on the final-state tracks.
For the kaon and proton in the $\Lbpk$ decay, 
the sum of the kaon and proton $\pt$ should be larger than $1\gevc$. 
Each \Kstarzb candidate is required to have a good vertex fit $\chisq$ and to have \pt greater than $1\gevc$.
The invariant mass of the reconstructed $\Kstarzb$ is required to be
within $\pm70\mevcc$ of the $\Kstarzb$ mass~\cite{PDG2014}.
Each $b$ hadron candidate must have a good vertex fit $\chisq$,
be consistent with originating from the PV,
and have a decay time greater than $0.2\ps$.

Some non-combinatorial backgrounds exist in the $\Lbpk$ data sample, 
originating from $\Bdb\to\jpsi\Km\pip$ and $\Bsb\to\jpsi\Km\Kp$ decays 
with the $\pip$ and $\Kp$ misidentified as a proton. 
In order to suppress these events, the invariant mass is recalculated
by interpreting the proton candidate as a pion or a kaon,
and the two relevant invariant mass regions are vetoed:
$m(\jpsi\Km\pip)\in[5250,5310]\mevcc$ and $m(\jpsi\Kp\Km)\in[5340,5400]\mevcc$. 
After the mass vetoes these background contributions are reduced to a negligible level.

After the preselection, the $\Lbpk$ ($\Bpik$) candidates are filtered with 
the BDT to further suppress combinatorial background.
For the decays $\Lbpk$ and $\Bpik$, the same BDT classifier is applied. 
Independent BDT classifiers are used for the 2011 and 2012 samples. 
In the BDT training a simulated $\Lb$ sample is used as the signal. 
The background is taken from the lower, $(5420,5560)\mevcc$,
and upper, $(5680,5820)\mevcc$, sidebands of the \Lb invariant mass distribution in data.
Events in the sidebands are randomly divided into two parts, one for the training
and the other for the test. No overtraining is observed.
The following information is used by the BDT classifier:
the kinematic properties and the impact parameters of the tracks;
and the vertex quality, the decay length and the impact parameter of the reconstructed $b$ hadron candidate.
The variables used for the training are chosen based on their power to discriminate 
signal from background and on the similarity of their distributions for 
$\Lbpk$ and $\Bpik$ decays. 
The threshold for the BDT response is chosen to maximise $S/\sqrt{S+B}$, 
where $B$ represents the number of background events 
estimated from the sideband region and $S$ the number of signal events in the mass peak.

%% file: X_determination.tex
\section{Cross-section and branching fraction determination}
\label{sec:X_determination}

The product of the differential production cross-section of each $b$ hadron 
and the corresponding branching fraction is calculated as
\begin{equation}
 {\deriv^2\sigma\over \deriv\pt\,\dy}~\BF
 =\frac{N(\pt,\y)}{\eps(\pt,\y)~\lum~\BR_{\inter}~\Dpt~\Dy},
\end{equation}
where $N(\pt,\y)$ and $\eps(\pt,\y)$ are respectively the signal yield
and the efficiency as functions of \pt and \y of the $b$ hadron,
$\Dpt$ and $\Dy$ are the bin widths,
$\lum$ is the integrated luminosity,
$\BF$ is the absolute branching fraction of the $\Lbpk$ ($\Bpik$) decay,
and $\BR_{\inter}$ represents the branching fractions of the intermediate decays:
\begin{equation*}
\BR_{\inter}\equiv
\begin{cases}
   \BR(\Jpsimumu) \qquad{\kern 7.8em} \mbox{for \Lb},\\
   \BR(\Jpsimumu)~\BR(\KstarzbKpi) \qquad \mbox{for \Bdb}.
   \end{cases}
\end{equation*}

The luminosity is measured with van der Meer scans 
and a beam-gas imaging method~\cite{LHCb-PAPER-2014-047}. 
The 2011 and 2012 data samples correspond to 
$1019\pm17\invpb$ and $2056\pm23\invpb$, respectively. 
The branching fraction $\BR(\jpsi\to\mumu)=(5.961\pm0.033)\%$~\cite{PDG2014},
while $\BR(\KstarzbKpi)$ is taken to be $2/3$ assuming isospin symmetry.
The branching fraction $\BR(\Bpik)=(1.29\pm0.05\pm0.13)\times10^{-3}$
as measured by \belle~\cite{Abe:2002haa} is used 
in preference to the world average value,
since in the \belle result the S-wave component is subtracted.

\section{Signal determination}
The signal yields of the $\Lbpk$ and $\Bpik$ decays are determined from 
unbinned extended maximum likelihood fits to the invariant mass distributions 
of the reconstructed $b$ hadron candidates 
in each $\pt$ and $y$ bin.  
In order to improve the mass resolution, 
the $b$ hadron is refitted with constraints~\cite{Hulsbergen:2005pu}
that it originates from the PV and that the reconstructed $\jpsi$ mass equals 
its known mass~\cite{PDG2014}.

\begin{figure}[tb]
\begin{center}
\includegraphics[width=0.48\linewidth]{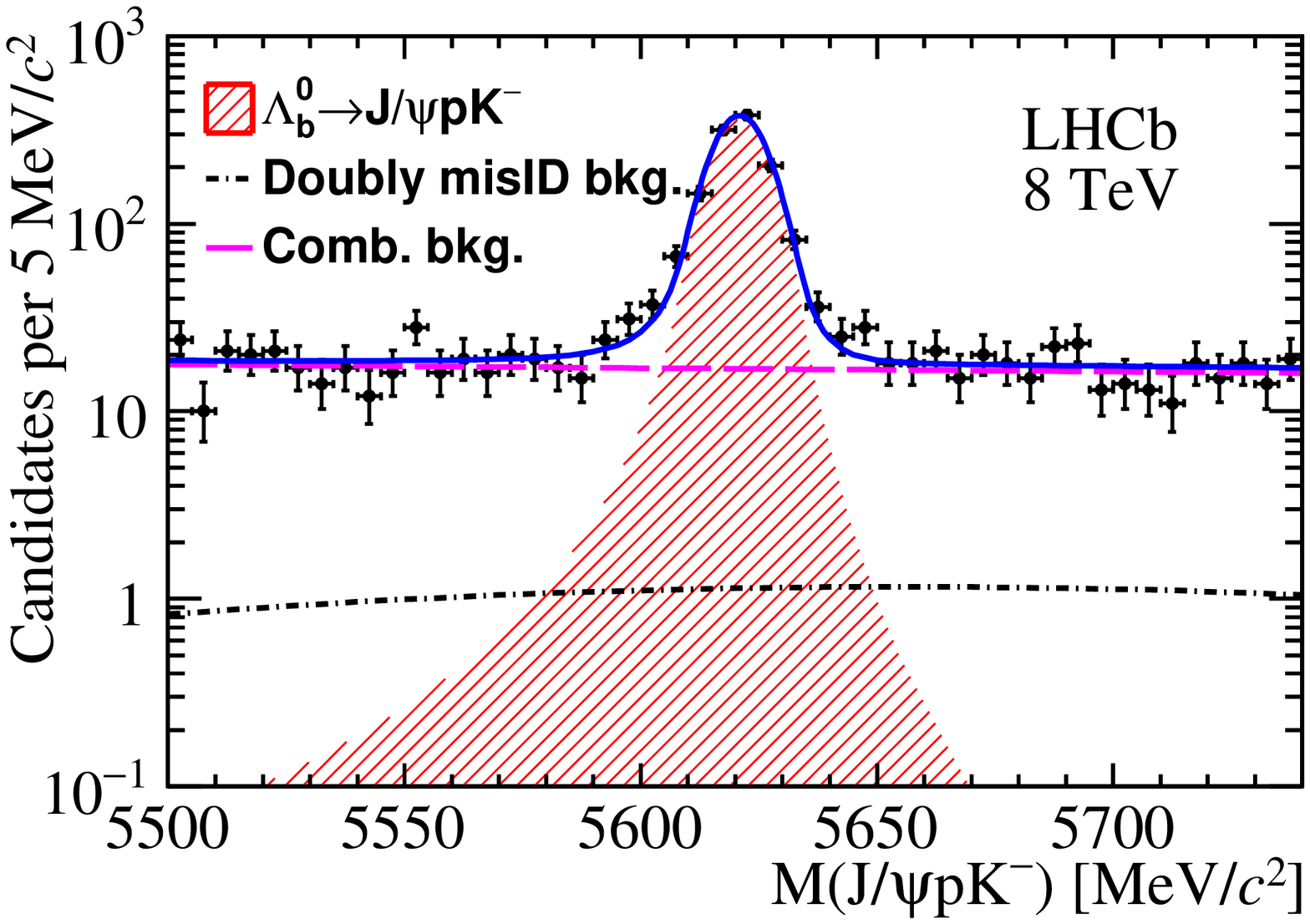}
\includegraphics[width=0.48\linewidth]{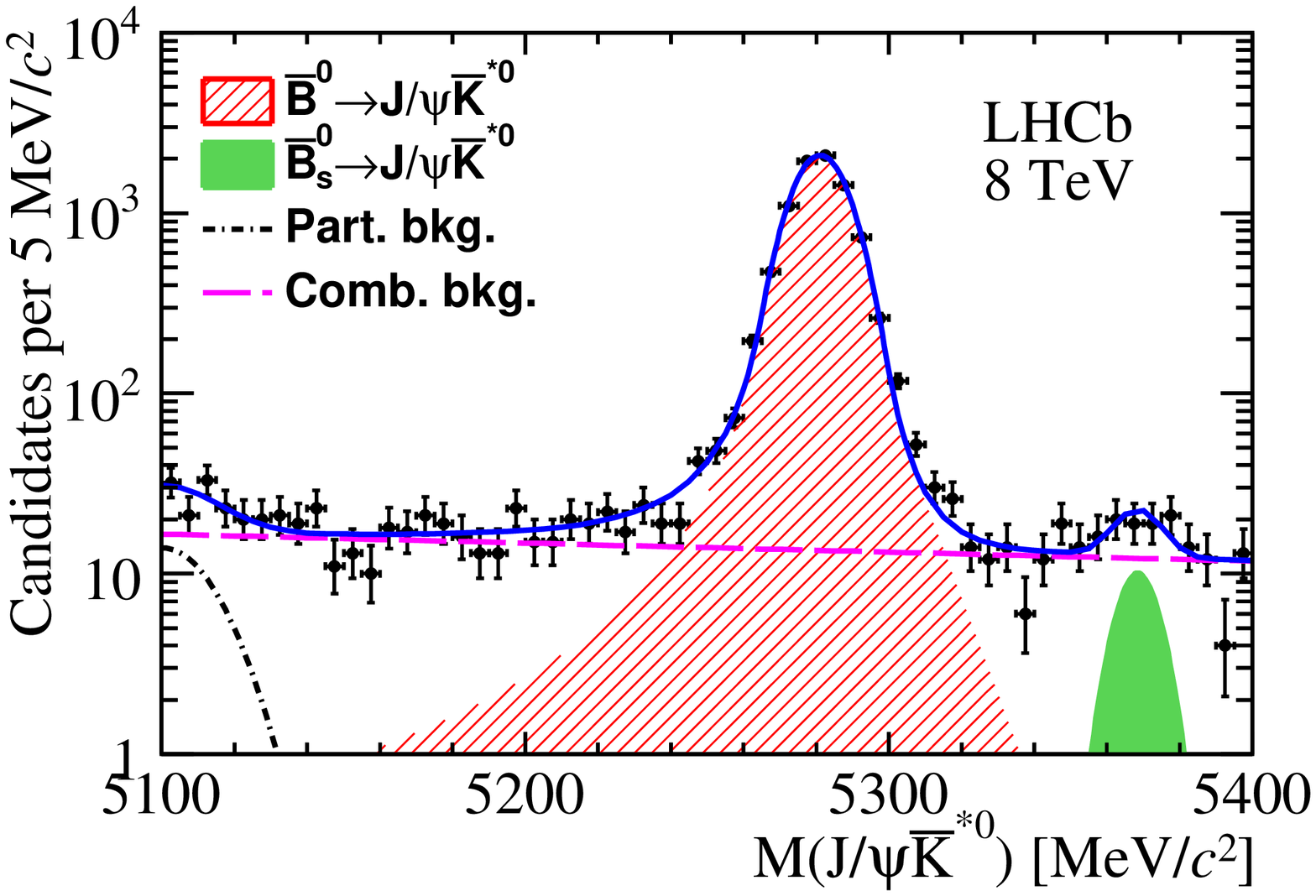}
\vspace*{-0.8cm}
\end{center}
\caption{\small
 Fit to the (left) $\jpsi\proton\Km$ and (right) $\jpsi\Kstarzb$ invariant mass 
 distributions with $\pt\in[6,7]\gevc$ and $y\in[3.0,3.5]$ for the 2012 data sample. 
 The hatched (red) area represents the signals, 
 the filled (green) area $\Bs\to\jpsi\Kstarzb$, 
 and the dashed (magenta) lines the combinatorial background. 
 The dot-dashed (black) lines indicate the doubly misidentified background (left) and partially reconstructed background (right). 
 The solid (blue) lines represent the sum of the above components and the points with error bars show the data.
}
\label{fig:signal_yield_bin}
\end{figure}

Figure~\ref{fig:signal_yield_bin} shows, as an example of one of the fit results, the invariant mass distributions
of $\Lb$ and $\Bdb$ candidates in the kinematic region $\pt\in[6,7]\gevc$ and $y\in[3.0,3.5]$ for the 2012 data sample.
The signal shape in the fits is modelled by a double-sided Crystal Ball (DSCB) function, 
an empirical function comprising a Gaussian core together with power-law tails on both sides.
The mean and the mass resolution of the DSCB function
are free in the fits, while the tail parameters are determined from simulation
in each kinematic bin according to the empirical function given in Ref.~\cite{LHCb-PAPER-2013-004}.
The combinatorial background is modelled by an exponential function whose parameters are left free in the fits. 

In the fits to the $\Lb$ mass distribution, there is a contribution 
from the $\Lbpk$ decay in which the proton is misidentified as a kaon 
and the kaon is misidentified as a proton.
This background is denoted as the doubly misidentified background, 
and it is modelled by a DSCB function. 
All parameters of this DSCB function are fixed from the simulation study, including:
the difference between the mean of this DSCB function and that of the signal shape;
the ratio of the mass resolution between these two DSCB; 
the yield fraction relative to the $\Lb$ signal channel;
and the tail parameters.

In the $\Bpik$ sample, in addition to the combinatorial background,
there are two further sources of background.
One is the decay $\Bs\to\jpsi\Kstarzb$, which populates the upper sideband of 
the invariant mass distribution, and is modelled with a DSCB function. 
The tail parameters of this DSCB function are the same as those of 
the $\Bdb$ signal shape and the remaining parameters are free in the fits.
The other comes from partially reconstructed $B$ mesons
and is described by the tail of a Gaussian function. 
The associated mean and width are free parameters in the fits.

According to a previous \lhcb measurement~\cite{LHCb-PAPER-2013-023}, 
the fraction of the $\Kstarzb$ meson contribution in the $\Bpik$ decay is calculated as $(89.9\pm0.4\pm1.3)\%$,
where the remainder is due to the S-wave component in the $\Km\pip$ system and its interference with the $\Kstarzb$ meson.
The fitted $\Bdb$ yields are subtracted with this number to remove the components from S-wave and its interference.   

\section{Efficiencies}
The efficiency $\eps^{\Lb,\Bdb}(\pt,y)$ consists of the geometrical acceptance 
of the detector, the trigger efficiency, the reconstruction and preselection efficiency,
the hadron PID efficiency, and the BDT selection efficiency. 
All the efficiencies are determined from a sample of simulated signal events, 
except the hadron PID efficiency, which is determined from data with tracks 
from the decays $\Jpsimumu$, $\Dstarp\to\Dz(\to\Km\pip)\pip$ and $\Lc\to\proton\Km\pip$.

The rich resonance structure observed in decays of \Lbpk in data~\cite{LHCb-PAPER-2015-029} is not modelled in the simulation.
The simulated sample  
is weighted to reproduce the distributions of the BDT training variables and the two-dimensional distribution of $m(p\Km)$ and $m(\jpsi\proton)$ observed in the background-subtracted data sample,
which has been obtained using the \sPlot technique~\cite{Pivk:2004ty},
with the $b$-hadron invariant mass as the discriminating variable.
It is found that the correlations between the discriminating variable and the control variables are negligible.

%% file: A_determination.tex
\section{Asymmetry determination}
\label{sec:A_determination}
The observed (raw) asymmetry for \Lb and \Lbbar 
is defined as
\begin{equation}
\label{eq:ArawDef}
 \Araw(x)\equiv\frac{N^{\Lb}(x)-N^{\Lbbar}(x)}{N^{\Lb}(x)+N^{\Lbbar}(x)}.
\end{equation}
The symbol $N(x)$ is the signal yield in the given bin of $x$ from the fits 
to the invariant mass distribution of the \Lb (\Lbbar) sample,
where $x$ denotes \pt or \y.
The observed asymmetry is a sum of several contributions:
the asymmetry between the numbers of the produced \Lb and \Lbbar baryons
in $pp$ collisions, $\aprod(x)$;
the decay asymmetry between the \Lbpk and \antiLbpk channels, $\adecay(x)$;
the asymmetry between the \proton and \antiproton detection efficiencies, $\aDproton(x)$;
the asymmetry between the \Km and \Kp detection efficiencies, $\aDKaon(x)$;
and the asymmetry between the PID efficiencies for \Lb and \Lbbar baryons, $\aPID(x)$.
Other possible asymmetries are neglected. 
Assuming that all these asymmetries are small, 
the asymmetries $\apd(x)$ of \Lb and \Lbbar baryons can be calculated as
\begin{equation}
\label{eq:apd}
\apd(x)=\Araw(x)-\aPID(x)-\aDproton(x)-\aDKaon(x). 
\end{equation}
The value $\Araw(x)-\aPID(x)$ can be calculated as
\begin{equation}
\Araw(x)-\aPID(x)=
  \frac{N^{\Lb}(x)/\eps^{\Lb}_{\rm PID}(x)-N^{\Lbbar}(x)/\eps^{\Lbbar}_{\rm PID}(x)}
       {N^{\Lb}(x)/\eps^{\Lb}_{\rm PID}(x)+N^{\Lbbar}(x)/\eps^{\Lbbar}_{\rm PID}(x)},
\end{equation}
where $\eps_{\rm PID}^{\Lb}(x)$ and $\eps_{\rm PID}^{\Lbbar}(x)$ represent the 
PID efficiencies for \Lb and \Lbbar.
The kaon detection asymmetry $\aDKaon(x)$ as a function of \pt and \y 
is obtained from a previous \lhcb study~\cite{LHCb-PAPER-2014-013}. 
The proton detection asymmetry $\aDproton(x)$ as a function of \pt or $y$ 
is estimated from simulation,
which uses the \geant model as described in Section~\ref{sec:Detector}.
The proton detection asymmetry as a function of \pt or \y is calculated with the proton 
and antiproton track reconstruction efficiencies in the corresponding kinematic bin. 
It is checked that the kinematic distributions of protons and \Lb baryons in the simulation sample are consistent with those in the data sample.
As a crosscheck, the proton detection asymmetries are also estimated through a simulation sample, where the \Lb signals are partially reconstructed without using the proton information, and the results are consistent.

%% file: systematics.tex
\section{Systematic uncertainties}
\label{sec:systematics}

\renewcommand\arraystretch{1.2}
\begin{table}[tb]
\caption{
 Summary of the systematic uncertainties (\%) for the production cross-sections 
 of $\Lb$ and $\Bdb$. The large uncertainties affect the bins with very few candidates.
}
\label{tab:systematics_x}
\vspace*{-0.5cm}
\begin{center}
\begin{tabular}{lcccc}
\hline
& $\Lb$ ($7\tev$) & $\Lb$ ($8\tev$) & $\Bdb$ ($7\tev$) & $\Bdb$ ($8\tev$)\\
\hline
{\it Uncorrelated between bins} & & & &\\
Signal shape         & $0.4-15.4  $ & $0.2-\xx6.2$ & $0.2-1.5$ & $0.2-\xx1.5$ \\
Background shape     & $0.0-\xx1.9$ & $0.0-\xx4.3$ & $0.0-0.9$ & $0.0-\xx0.9$ \\
Simulation sample size & $4.1-16.5  $ & $3.9-14.3  $ & $1.7-9.5$ & $2.2-14.9  $ \\
BDT efficiency       & $0.4-\xx2.5$ & $0.4-\xx2.8$ & $0.1-0.5$ & $0.1-\xx0.5$ \\
Trigger efficiency   & $0.0-\xx4.6$ & $0.0-14.9  $ & $0.0-2.1$ & $0.0-\xx4.0$ \\
PID efficiency       & $0.4-\xx8.4$ & $0.4-15.8  $ & $0.2-4.6$ & $0.2-\xx2.7$ \\
Resonance            &  &  & $0.0-1.0$ & $0.0-\xx1.8$  \\
\hline
{\it Correlated between bins}& &      &       &\\
Tracking efficiency   & $3.0$ & $3.0$ & $3.0$ & $3.0$ \\
Mass veto efficiency  & $1.3$ & $1.9$ &       &       \\
Luminosity            & $1.7$ & $1.2$ & $1.7$ & $1.2$ \\
$\BR(\jpsi\to\mumu)$  & $0.6$ & $0.6$ & $0.6$ & $0.6$ \\
S-wave and interference in $\Km\pip$   &       &       & $1.4$ & $1.4$ \\
\hline
\end{tabular}
\end{center}
\end{table}
\begin{table}[tb]
\caption{
Summary of the absolute systematic uncertainties (\%) for the asymmetry of $\Lb$ and $\Lbbar$.
The large uncertainties affect the bins with very few candidates.
}
\label{tab:systematics_a}
\vspace*{-0.5cm}
\begin{center}
\begin{tabular}{lcc}
\hline
& 2011 & 2012 \\
\hline
PID efficiency               & $0.4-4.4$ & $0.0-2.6$ \\
Signal shape                 & $0.0-0.8$ & $0.0-0.9$ \\
Background shape             & $0.0-0.1$ & $0.0-0.3$ \\
MC statistics                & $0.7-5.4  $ & $0.3-4.2$ \\
Tracking asymmetry of proton & $0.1-1.9$ & $0.1-1.9$ \\
\hline
\end{tabular}
\end{center}
\end{table}

Several sources of systematic uncertainties are studied in the analysis
and are summarised in Tables~\ref{tab:systematics_x} and \ref{tab:systematics_a}.
For the production cross-section measurements, the uncertainties originate from
the determination of the signal yields, efficiencies, branching fractions and luminosities.
The total systematic uncertainties are obtained from the sum in quadrature of all components.

Imperfect knowledge of the mass distributions for the signal and backgrounds 
causes systematic uncertainties in the signal yield determination.
For the signal shape, the Apollonios function~\cite{Santos:2013gra}
and the sum of a Gaussian function and a Crystal Ball function 
are tried as alternatives to the DSCB.
The largest deviation to the nominal result is taken as the uncertainty 
due to the model of the signal shape. 

The fits are repeated with a linear function substituted for the exponential model for the combinatorial background.
The fits are also repeated without the double misidentified components.
The maximum differences of the signal yields from the nominal results are quoted as systematic uncertainties due to the background shape.

Most efficiencies are estimated from simulation. 
The limited size of the simulation sample leads to systematic uncertainties on the efficiencies ranging from $1.7\%$ to $16.5\%$.

The tracking efficiency is estimated from simulation and calibrated by data~\cite{LHCb-DP-2013-002}. 
The uncertainty of the calibration is $0.4\%$ per track.
Additional systematic uncertainties are assigned to hadrons due to imperfect knowledge
of hadron interactions in the detector, $1.1\%$ for kaons, $1.4\%$ for pions and $1.4\%$ for protons.

The BDT efficiency is estimated with the weighted simulation sample
to ensure that the distributions of the two training variables, the kinematic properties of the tracks and the vertex quality, agree with those in data.
The uncertainties on the weights are propagated to the final results
to give the corresponding systematic uncertainty.

The trigger efficiency is determined in the simulation and validated 
in a control sample of $\Jpsimumu$ decays~\cite{LHCb-DP-2012-004}.
The difference of the central values of this determination in data and the simulation 
in each bin is taken as the systematic uncertainty.
Uncertainties due to the limited sample size of the simulation are added in quadrature.

The PID efficiency is estimated with a data-driven method.
A sample of $\Jpsimumu$, $\Dstarp\to\Dz(\to\Km\pip)\pip$ and $\Lc\to p\Km\pip$ 
decays obtained without using PID information is used to evaluate the PID efficiency.
The limited sample size used to calculate the PID efficiency 
introduces a systematic uncertainty in each kinematic bin.
To study the bin-by-bin migration effect, the number of the bins is doubled or halved
and the PID efficiency is recalculated.
The largest deviation from the nominal result is taken as the uncertainty.

To account for the rich and complex structure of multiple intermediate resonances 
in the \Lbpk decay, the simulation sample is weighted in two-dimensional bins of $m(\Km\proton)$ 
and $m(\jpsi\proton)$ to match the data. 
Pseudoexperiments are performed to estimate the systematic uncertainties due to the weights.
The weight in each bin is varied according to its uncertainty and the total efficiency is recalculated.
The RMS of the distribution obtained from 
the pseudoexperiments is taken as the systematic uncertainty.
As mentioned in Section~\ref{sec:selection}, the preselection includes mass vetoes. The preselection efficiencies are estimated from the simulation sample.
A fit to the \Lb invariant mass distribution in the vetoed data sample is performed, 
which gives the number of \Lbpk signal events rejected by the vetoes.
The fraction of the vetoed signal events in the data sample is compared with that in the simulation sample.
A difference of $1.3\%$ ($1.9\%$) is observed for the 2011 (2012) sample,
and this is taken as the systematic uncertainty.

The uncertainty in the determination of the integrated luminosity 
is $1.7\%$ ($1.2\%$) for the 2011 (2012) data sample~\cite{LHCb-PAPER-2014-047}.
An uncertainty of $0.6\%$ is taken on $\BF(\Jpsimumu)$~\cite{PDG2014}.
The fractions of the S-wave component in the $\Km\pip$ system and their interference were determined
by a previous LHCb measurement, and their $1.4\%$ uncertainty~\cite{LHCb-PAPER-2013-023}
is taken as a systematic uncertainty for the \Bpik decay.

In the $\Lb$ and $\Lbbar$ asymmetry  
measurement, 
all of the uncertainties mentioned above cancel in the ratio,
except for those due to the signal shape, the background shape, 
the limited sample size and the PID efficiency.
Since a data-driven determination of proton detection asymmetries is not available,
the difference in the determination of the kaon detection asymmetries 
in data and simulation is taken as a systematic uncertainty 
for the proton detection asymmetry.
The uncertainties vary from $0.1\%$ to $1.9\%$ in kinematic bins, with large values 
occurring in bins of low \pt or low signal yields.
In the \lhcb \geant physics models, the cross-sections of interactions between particles 
and the material are checked with test beam data as discussed in Section~\ref{sec:Detector}. There are more data for protons 
than for kaons. Therefore, these uncertainties can be considered to be conservative.

%% file: result.tex
\section{Cross-section results}
\label{sec:result}

\begin{figure}[tb]
\begin{center}
 \includegraphics[width=0.49\linewidth]{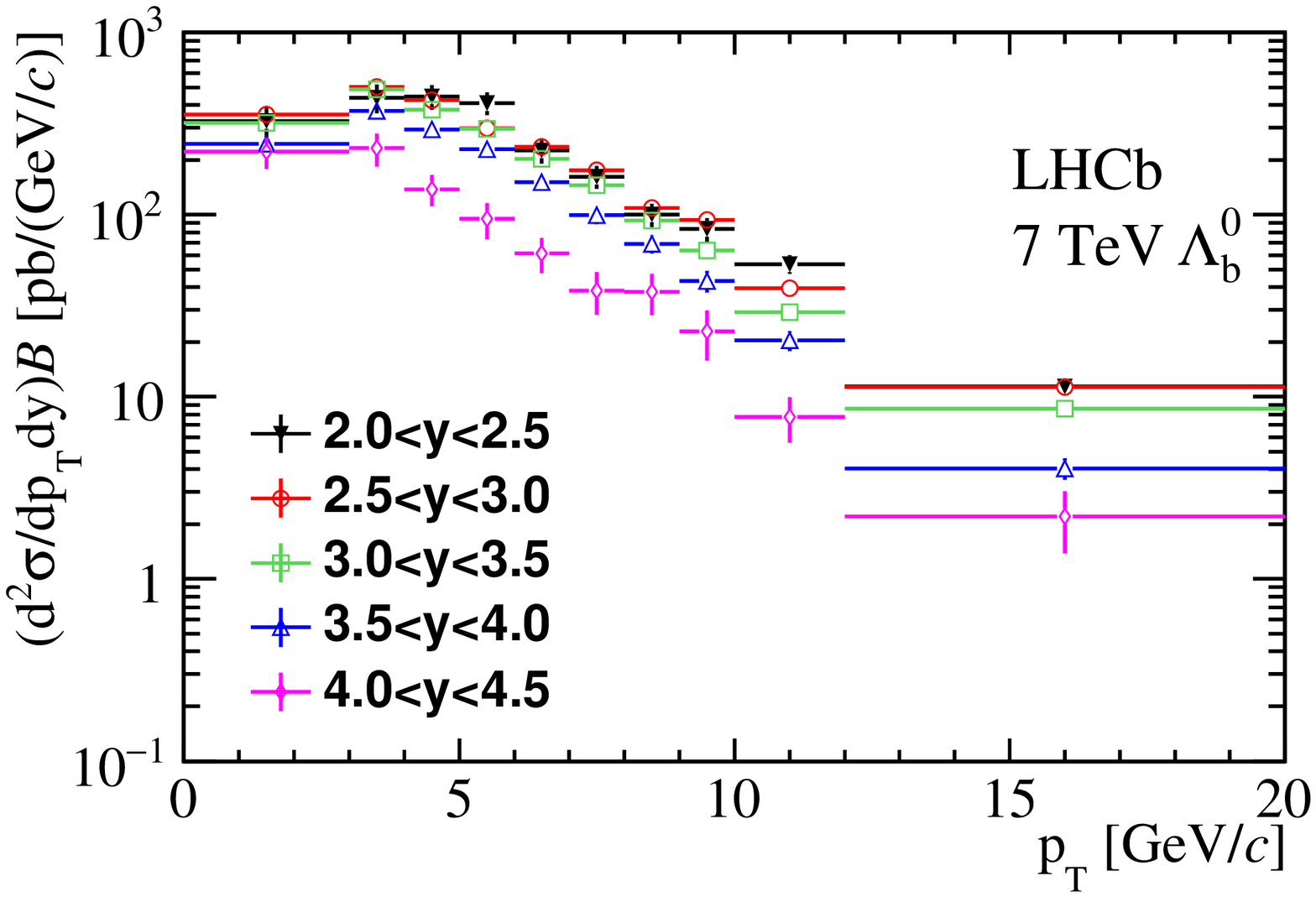}
 \includegraphics[width=0.49\linewidth]{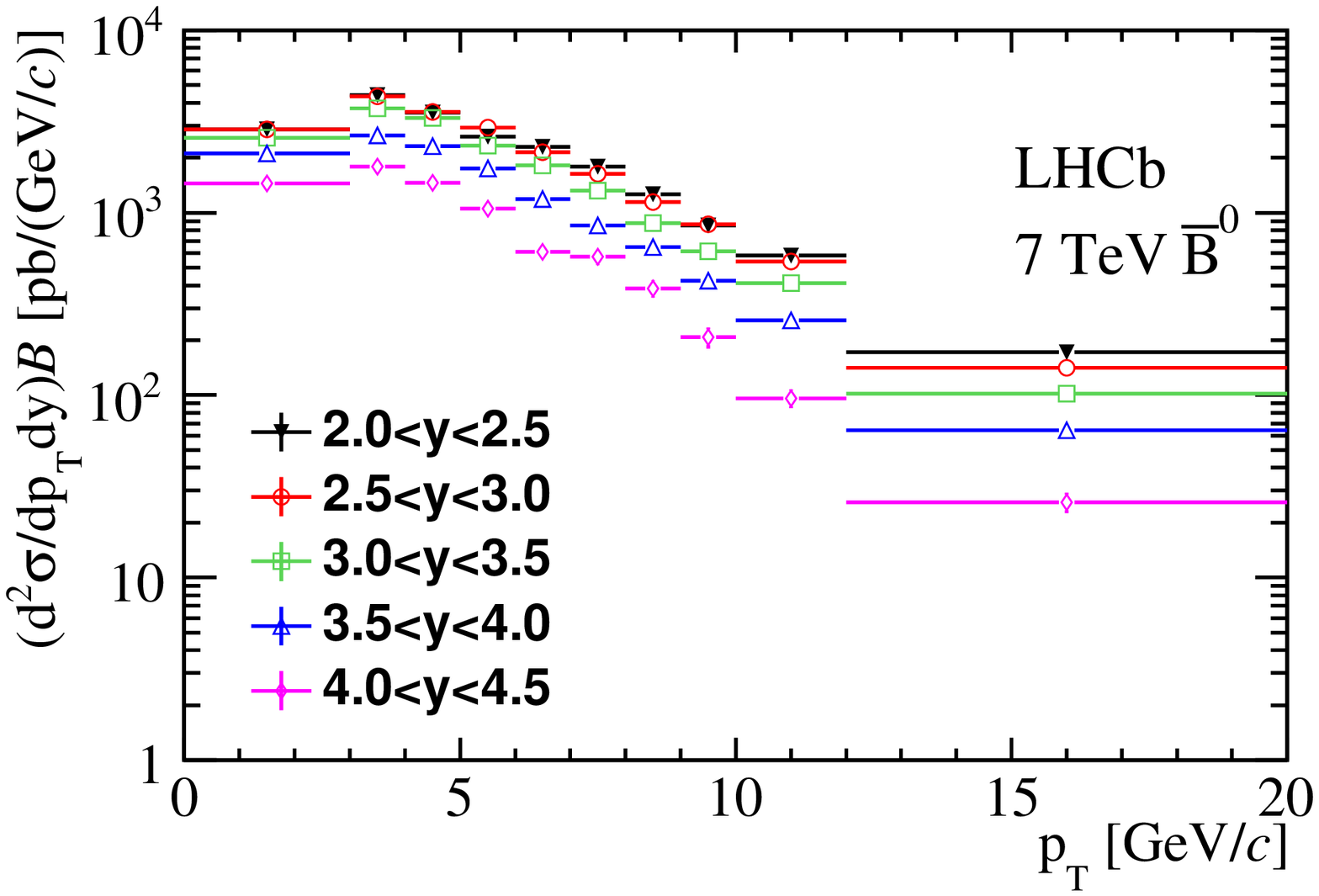}\\
 \includegraphics[width=0.49\linewidth]{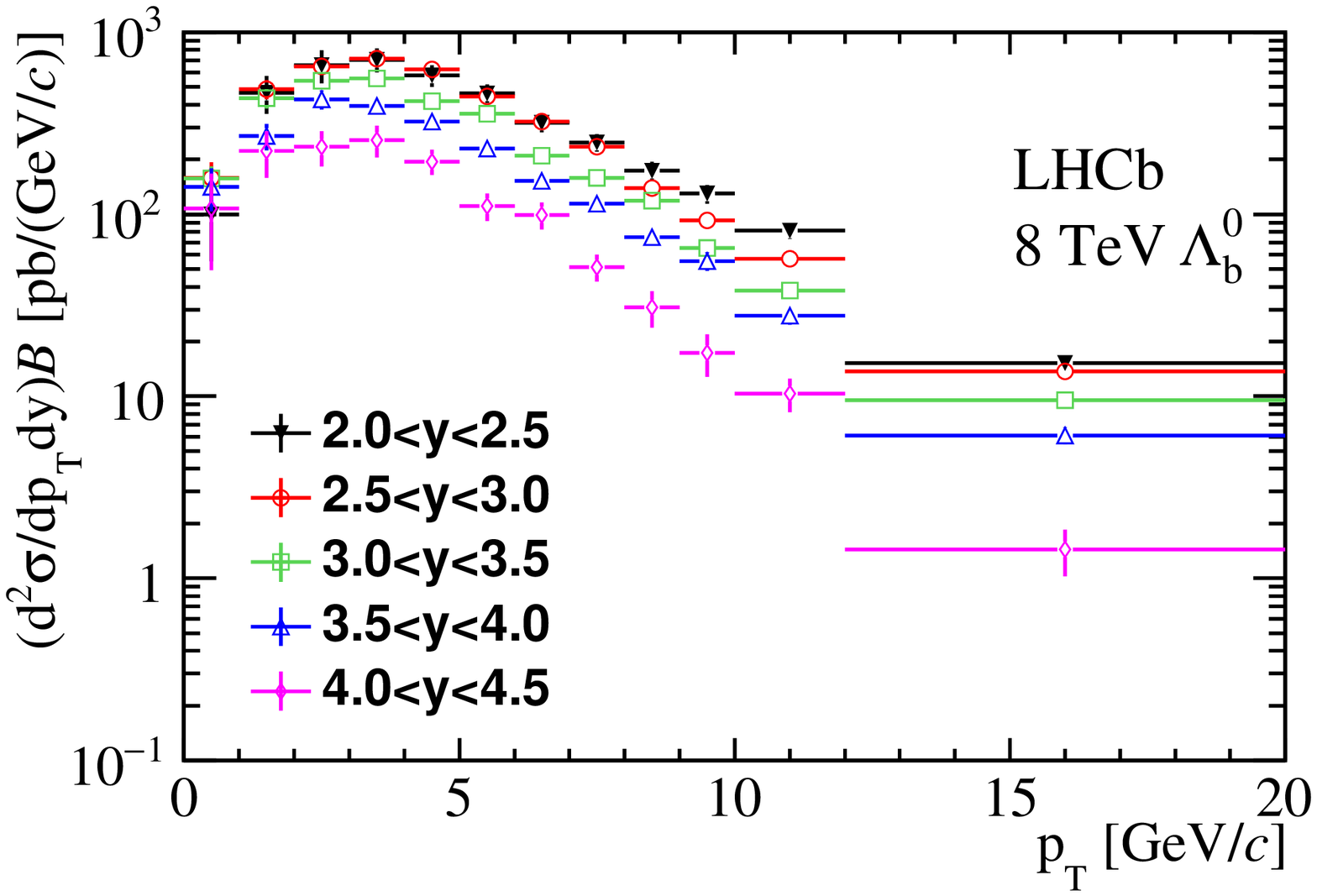}
 \includegraphics[width=0.49\linewidth]{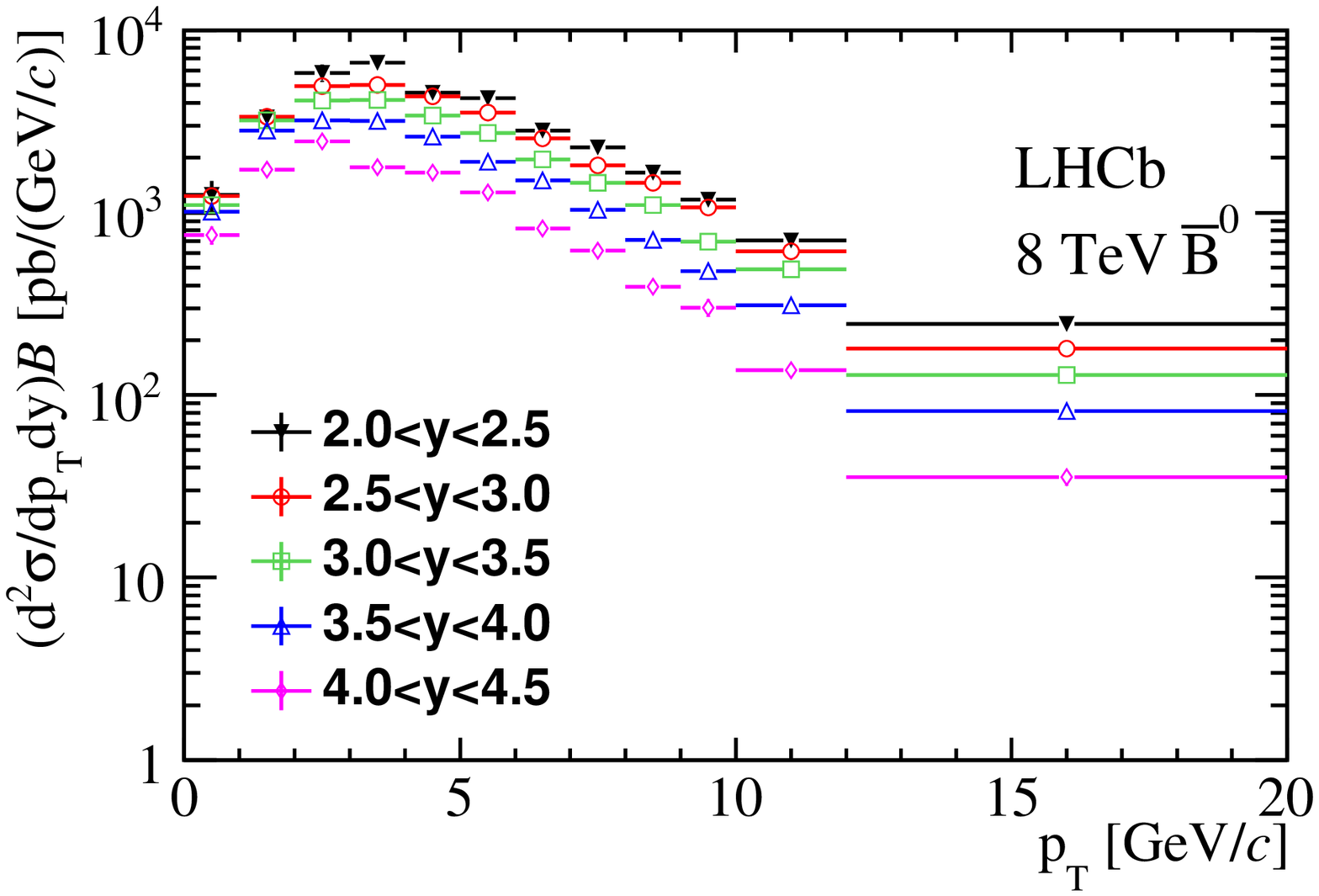}
\end{center}
\vspace*{-0.5cm}
\caption{
 Products of production cross-sections and branching fractions 
 as functions of \pt in $y$ bins for (left) \Lbpk and (right) \Bpik.
 The top (bottom) plots represent the 2011 (2012) sample.
 The error bars 
 represent the total uncertainties.
}\label{fig:xans}
\end{figure}

\begin{figure}[tb]
\begin{center}
 \includegraphics[width=0.49\linewidth]{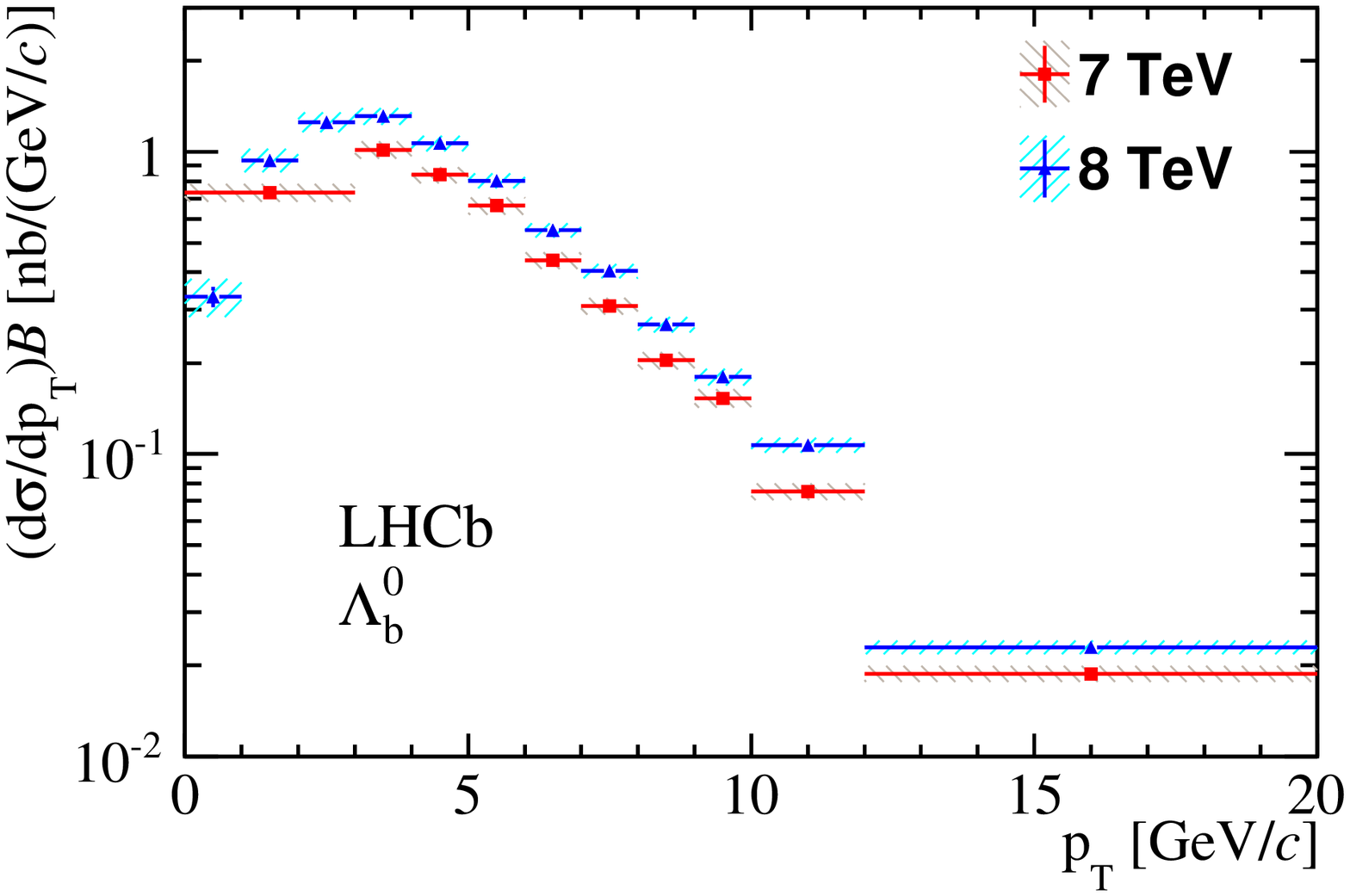}
 \includegraphics[width=0.49\linewidth]{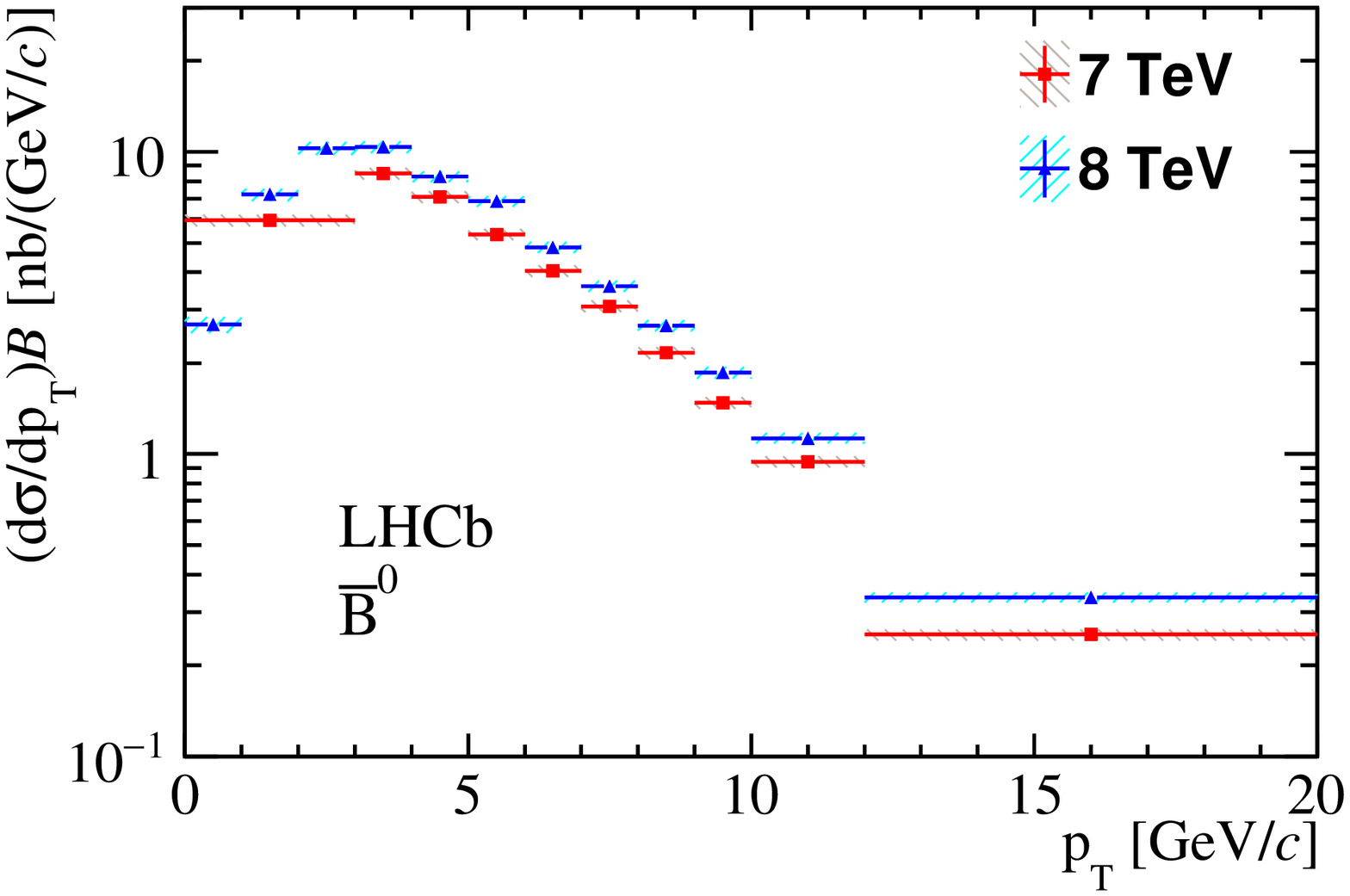}\\
 \includegraphics[width=0.49\linewidth]{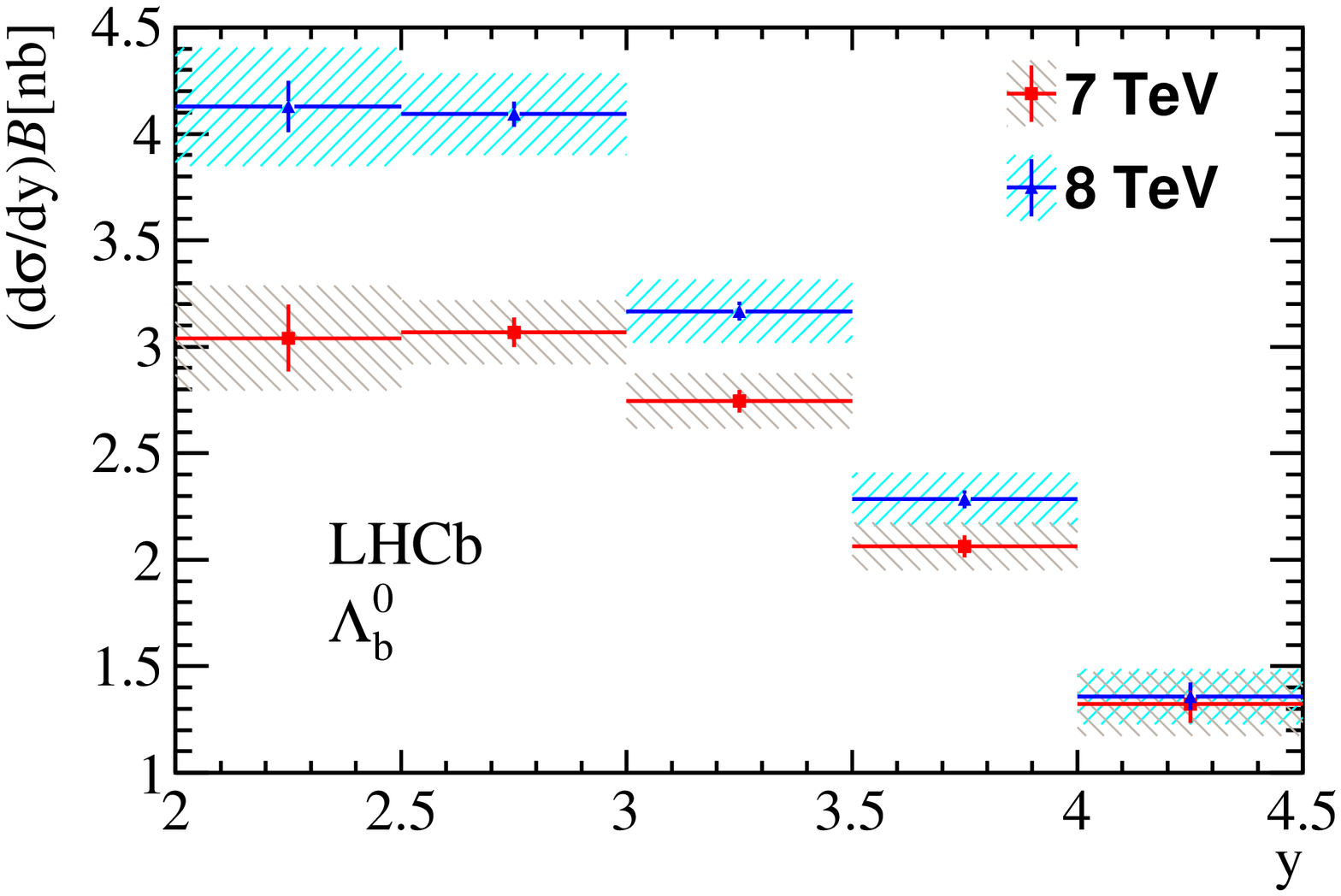}
 \includegraphics[width=0.49\linewidth]{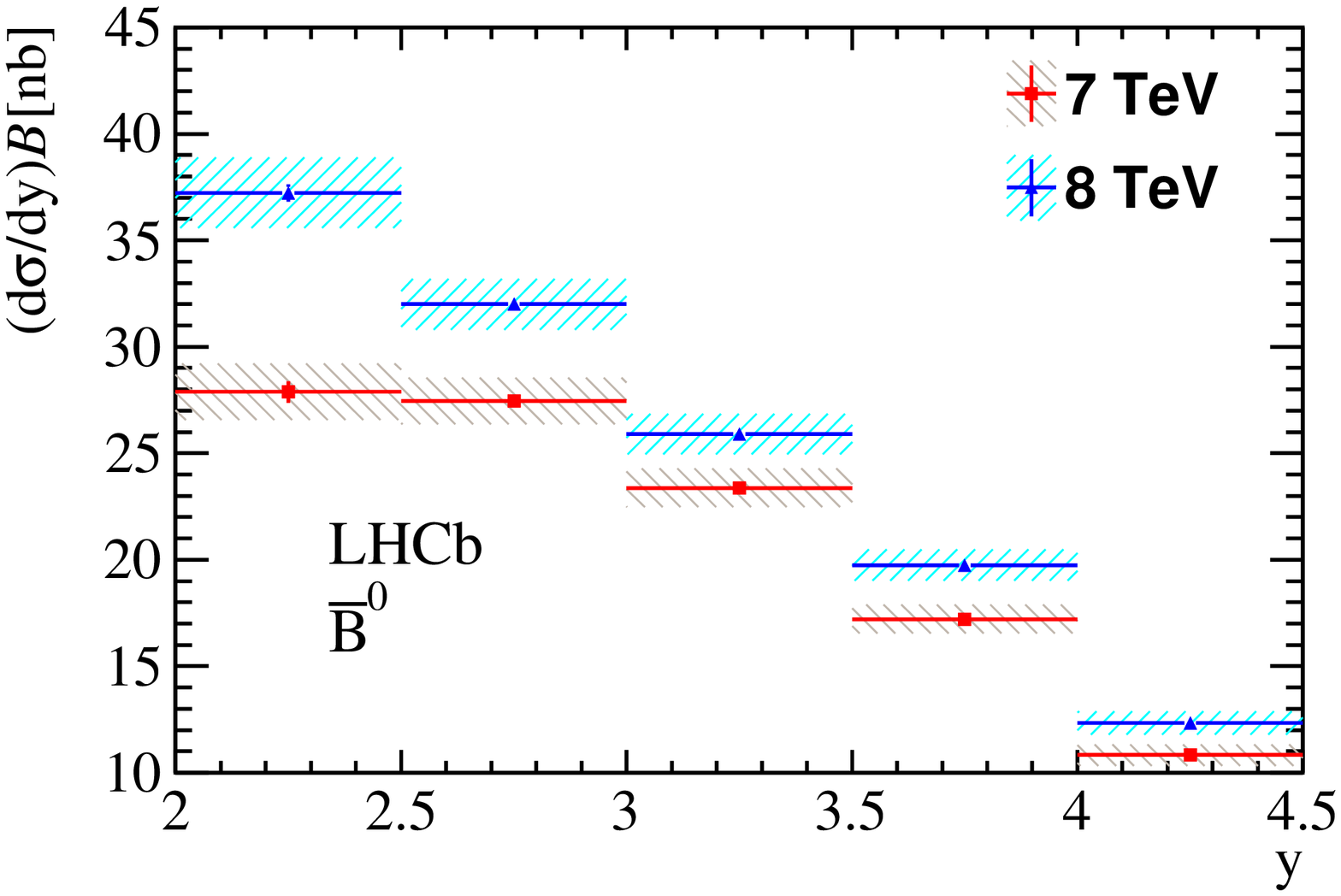}
\end{center}
\vspace*{-0.5cm}
\caption{ 
 Products of production cross-sections and branching fractions 
 as functions of (top) \pt or (bottom) \y. 
 The left (right) plots represent \Lb (\Bdb) hadrons. 
 The error bars indicate the statistical uncertainties 
 and the hatched areas represent the total uncertainties.
}
\label{fig:xans_1D}
\end{figure}

\begin{figure}[t]
\begin{center}
 \includegraphics[width=0.49\linewidth]{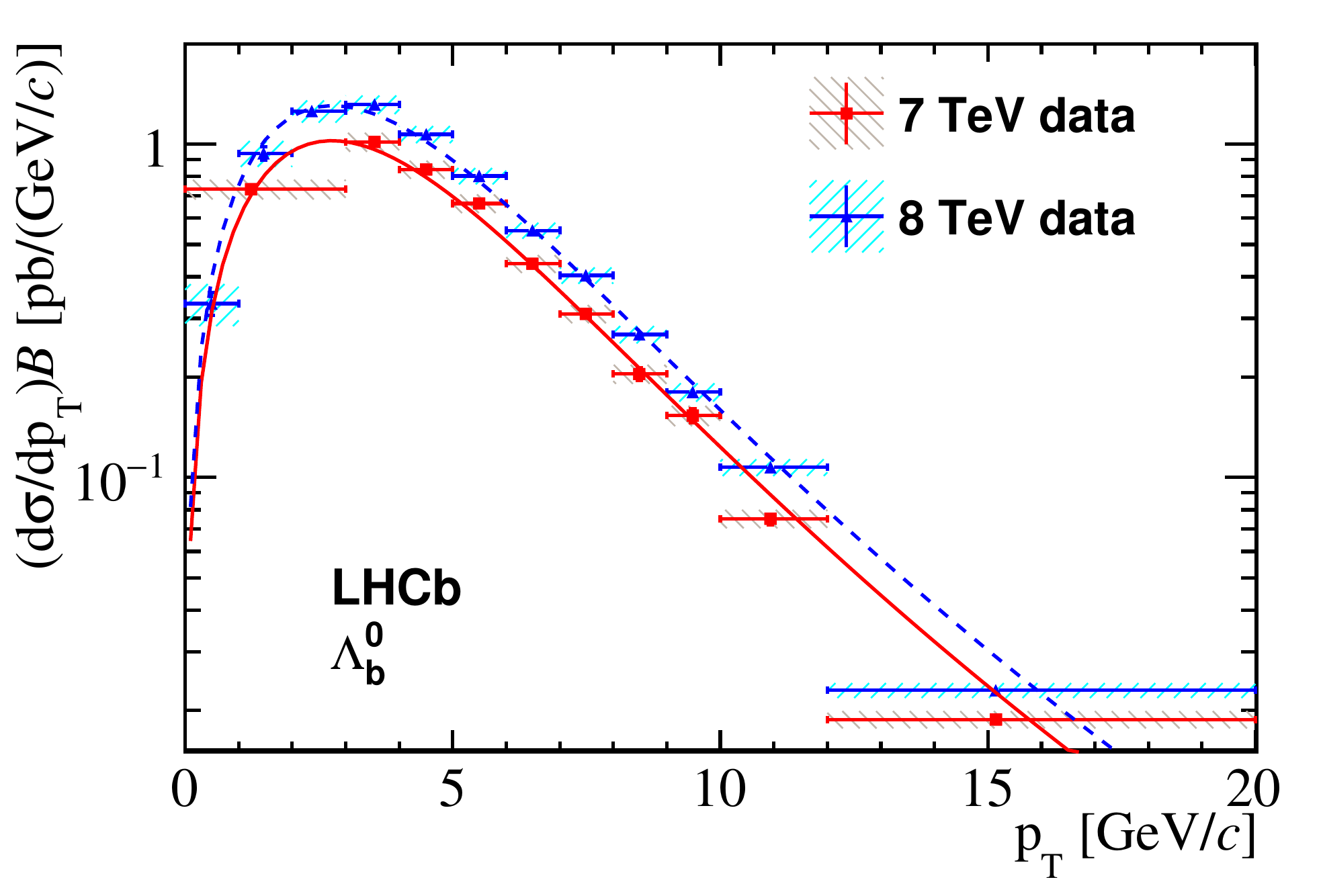}
\end{center}
\vspace*{-0.5cm}
\caption{
 Fit to the \Lb distribution with the Tsallis function.
}\label{fig:fitLb1Dpt}
\end{figure}

The product of the \Lb (\Bdb) double-differential cross-section 
and the branching fraction of the decay \Lbpk (\Bpik)
is shown in Fig.~\ref{fig:xans}, and the values are listed in 
Tables~\ref{tab:final_result_Lb11},~\ref{tab:final_result_Lb12},~\ref{tab:final_result_B011} and~\ref{tab:final_result_B012} in the Appendix. 
By integrating over \y or \pt, the single differential production 
cross-sections, shown in Fig.~\ref{fig:xans_1D}, are obtained.
Figure~\ref{fig:fitLb1Dpt} shows the \pt distribution of the $\Lb$ production,
fitted by a power-law function with the Tsallis parameterisation~\cite{Tsallis:1987eu,Tsallis:1999nq}:
\begin{equation}
 \frac{\deriv\sigma}{\pt\dpt}\propto\frac{1}{\left[1+E_{k\perp}/(TN)\right]^N},
\end{equation}
where $T$ is a temperature-like parameter, 
$N$ determines the power-law behaviour at large $E_{k\perp}$,
and $E_{k\perp}\equiv\sqrt{\pt^2+M^2}-M$ with $M$ the mass of the hadron. 
The fit results are 
\begin{equation*}
 \begin{aligned}
  &T=1.12\pm0.04~\gev\qquad N=7.3\pm 0.5\qquad (7\tev),\\
  &T=1.13\pm0.03~\gev\qquad N=7.5\pm 0.4\qquad (8\tev).
 \end{aligned}
\end{equation*}
For the $7\tev$ ($8\tev$) sample, the fit $\chi^2$ is 21.0 (10.7) 
for 7 (9) degrees of freedom.
The parameters $T$ and $N$ obtained from the $7\tev$ and $8\tev$ samples 
are consistent with each other and with the values found 
by CMS~\cite{Chatrchyan:2012xg}. 
Other functions suggested in Ref.~\cite{Bylinkin:2015gqa} do not give acceptable fits to the data.
In Fig.~\ref{fig:fitLb1Dpt} the data points are placed in the bin according to the prescription of Ref.~\cite{Lafferty:1994cj}.

\begin{figure}[tb]
\begin{center}
 \includegraphics[width=0.49\linewidth]{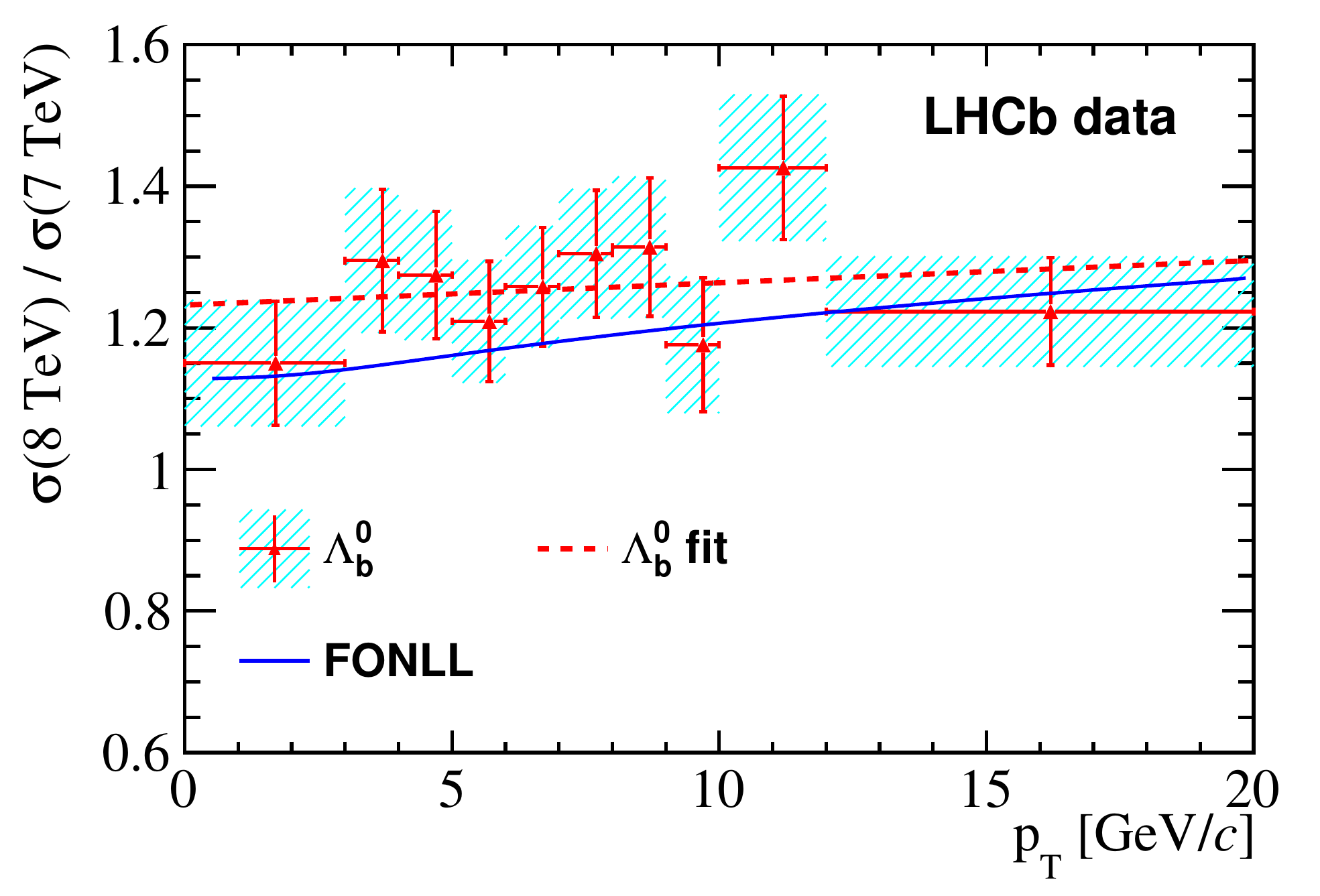}
 \includegraphics[width=0.49\linewidth]{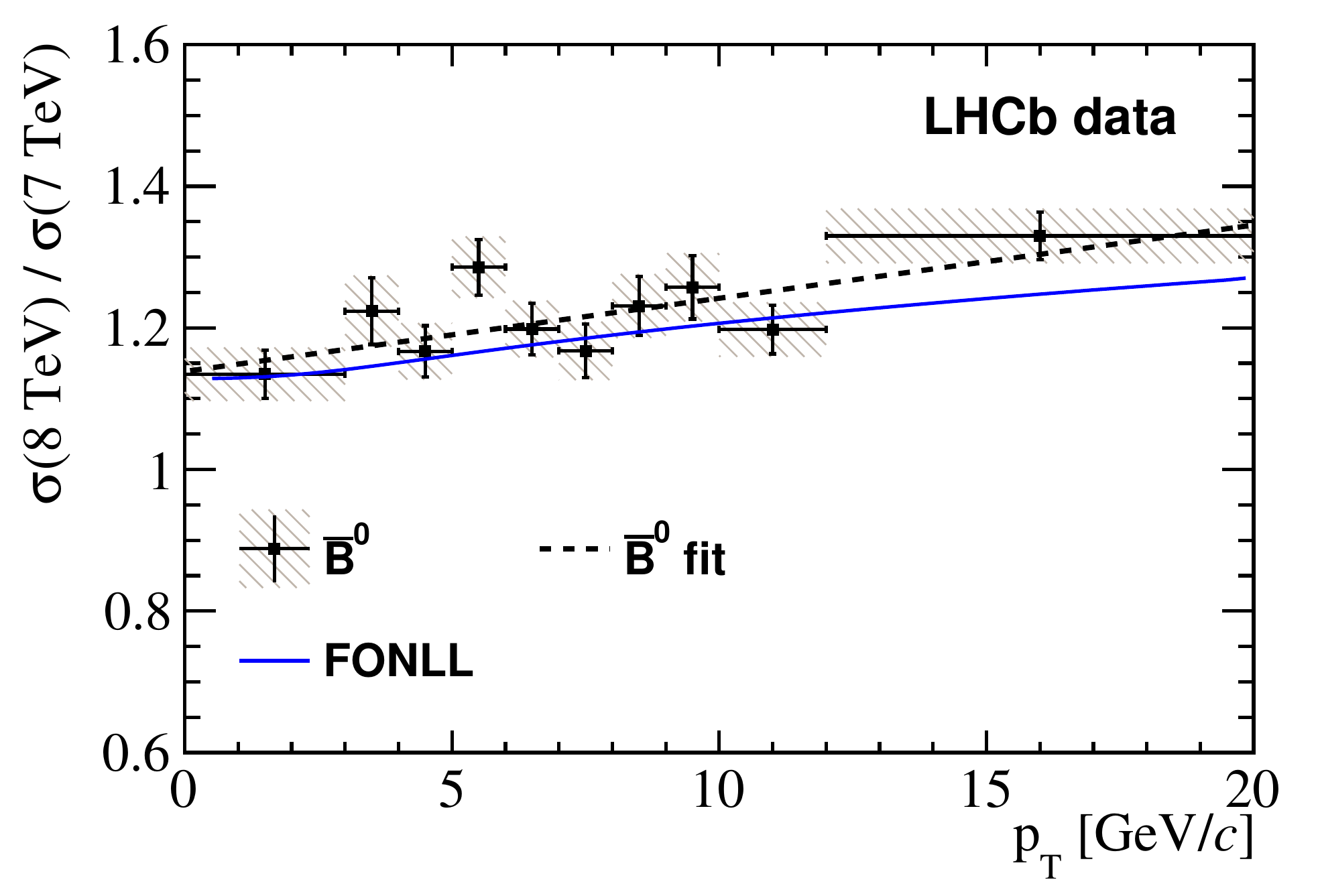}\\
 \includegraphics[width=0.49\linewidth]{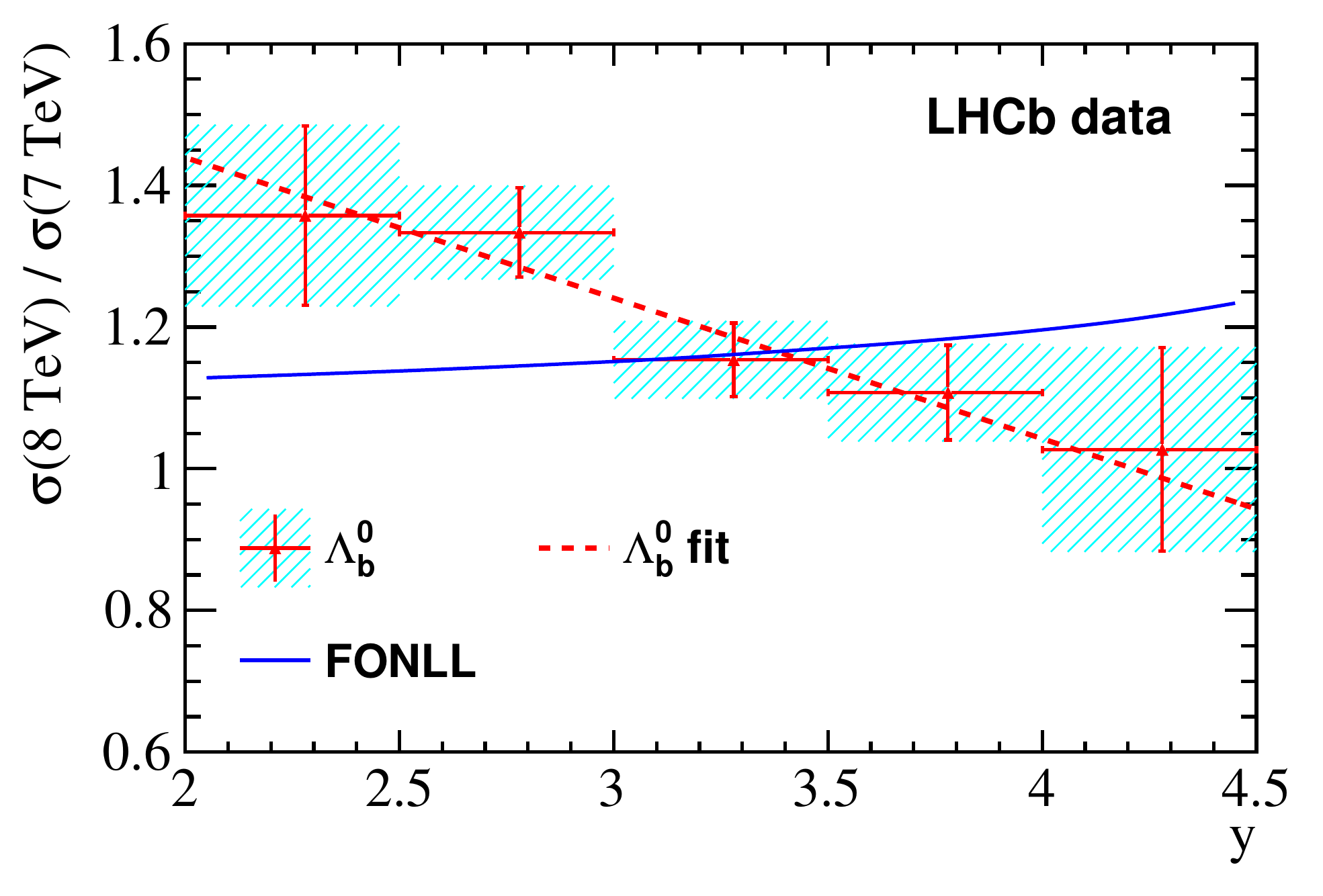}
 \includegraphics[width=0.49\linewidth]{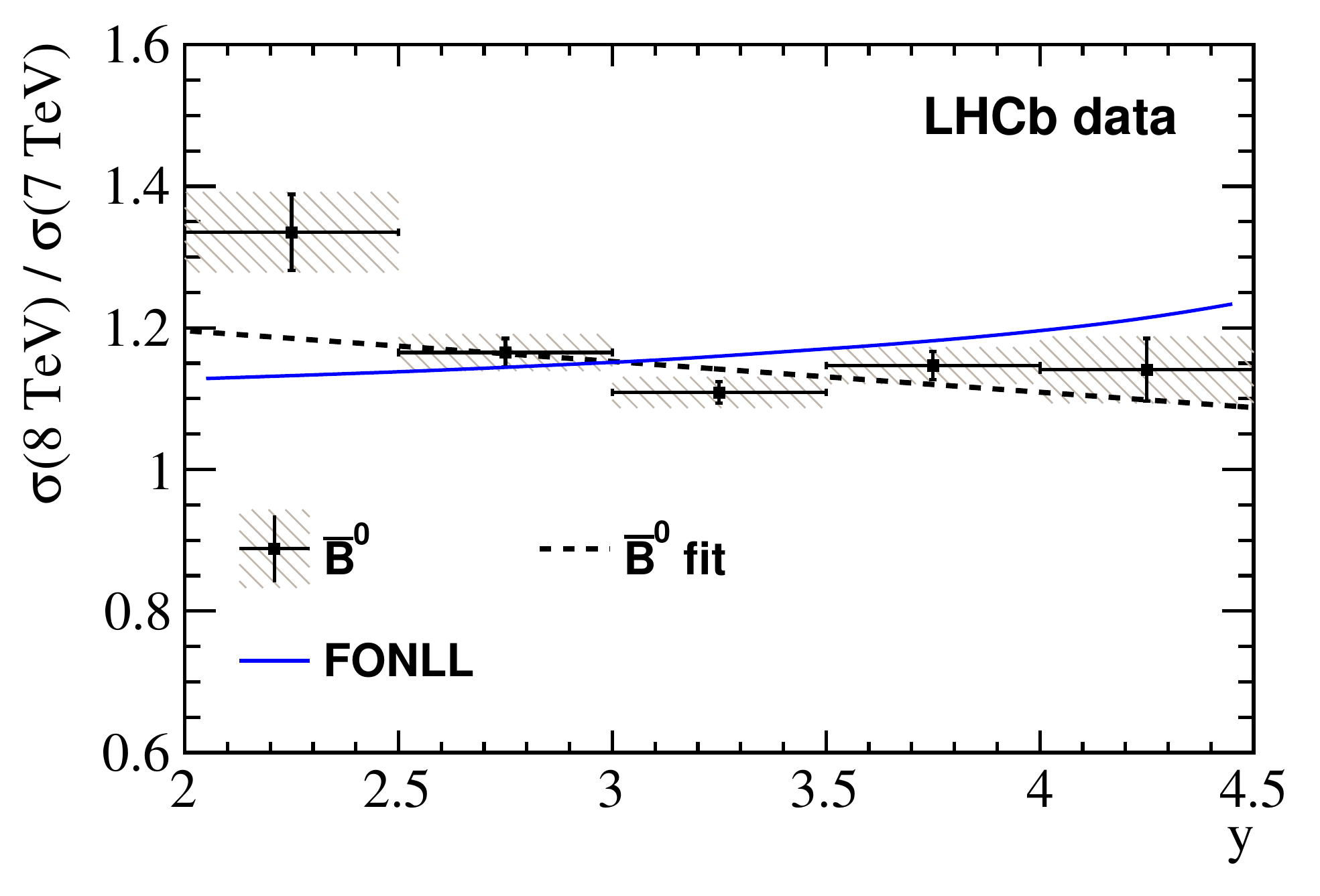}
  \vspace*{-0.5cm}
 \end{center}
 \caption{ 
 Production ratios of (left) $\Lb$ and (right) $\Bdb$ at $8\protect\tev$ and $7\protect\tev$
 as functions of the (top) $\pt$ and (bottom) $y$ of the $b$ hadron.
 The blue lines are FONLL predictions.
 The error bars represent uncorrelated uncertainties, 
 while the hatched areas show the total uncertainties.
 Linear fits are also shown.
 }
\label{fig:8and7ratio}
\end{figure}

The integrated cross-sections of the $b$ hadrons with $0<\pt<20\gevc$ and $2.0<y<4.5$ are measured to be
\begin{eqnarray*}
&&\sigma(\Lb,~\sqrt{s}=7\tev)~\BR(\Lbpk)\\
&=&\input{table/number/integral/Lb11/ans.tex},\\
&&\sigma(\Lb,~\sqrt{s}=8\tev)~\BR(\Lbpk)\\
&=&\input{table/number/integral/Lb12/ans.tex},\\
&&\sigma(\Bdb,~\sqrt{s}=7\tev)~\BR(\Bpik)\\
&=&\input{table/number/integral/B011/ans.tex},\\
&&\sigma(\Bdb,~\sqrt{s}=8\tev)~\BR(\Bpik)\\
&=&\input{table/number/integral/B012/ans.tex}.
\end{eqnarray*}
Taking the branching fraction $\BR(\Bpik)$ from \belle~\cite{Abe:2002haa}, 
the measured \Bdb production cross-section at $7\tev$ is consistent with 
the previous \lhcb measurement~\cite{LHCb-PAPER-2013-004}.
The ratios of the \Lb and \Bdb integrated production cross-sections 
between $8\tev$ and $7\tev$, in the kinematic range $0<\pt<20\gevc$ and $2.0<\y<4.5$, 
are
\begin{equation*}
 \frac{\sigma(\sqs=8\tev)}{\sigma(\sqs=7\tev)}=
 \begin{cases}
  &1.23\pm0.02\pm0.04\qquad\mbox{for }\Lb,\\
  &1.19\pm0.01\pm0.02\qquad\mbox{for }\Bdb,
 \end{cases}
\end{equation*}
where the first uncertainties are statistical and the second systematic.
Many systematic uncertainties cancel totally or partially in these ratios:
the ratio of the luminosities is known with a precision of $1.44\%$~\cite{LHCb-PAPER-2014-047};
the tracking efficiency is considered to be fully correlated, 
due to the fact that the associated systematic uncertainty 
is dominated by hadronic interactions of the tracks in the detector;
the mass veto efficiency, the branching fraction of the $\Jpsimumu$ decay,
and the S-wave contribution in the $\Km\pip$ system are also fully correlated.
All other sources are considered uncorrelated. 
The ratio of the integrated production cross-sections agrees with 
FONLL predictions~\cite{Cacciari:1998it,Cacciari:2001td,Cacciari:2012ny}.
Figure~\ref{fig:8and7ratio} shows the \pt and $y$ dependence of the ratios 
for \Lb and \Bdb production cross-sections at $8\tev$ with respect to those at $7\tev$,
together with linear fits to the distributions:
\begin{equation*}
 \begin{split}
 &\Lb 
\begin{cases}
 &(1.25\pm0.03) + (0.003\pm0.007)(\pt-\mean{\pt})/(\mygevc) \qquad \chisqndf=6.1/8,\\
 &(1.22\pm0.03) - (0.20\pm0.07)(y-\mean{y}) \hspace{3.36cm} \chisqndf=1.2/3,
\end{cases}\\
 &\Bdb 
\begin{cases}
&(1.21\pm0.01) + (0.010\pm0.003)(\pt-\mean{\pt})/(\mygevc) \qquad \chisqndf=12/8,\\
&(1.15\pm0.01) - (0.04\pm0.02)(y-\mean{y}) \hspace{3.36cm} \chisqndf=15/3,
\end{cases}
 \end{split}
\end{equation*}
where $\mean{\pt}=6.7~(6.9)\gevc$ is the mean \pt of \Lb (\Bdb) hadrons
in the data sample, $\mean{\y}=3.1$ is the mean \y, 
and $\ndf$ is the number of degrees of freedom.
The \pt dependence of the ratio agrees with FONLL predictions, 
while the $y$ dependence does not.

\begin{figure}[tb]
\begin{center}
 \includegraphics[width=0.49\linewidth]{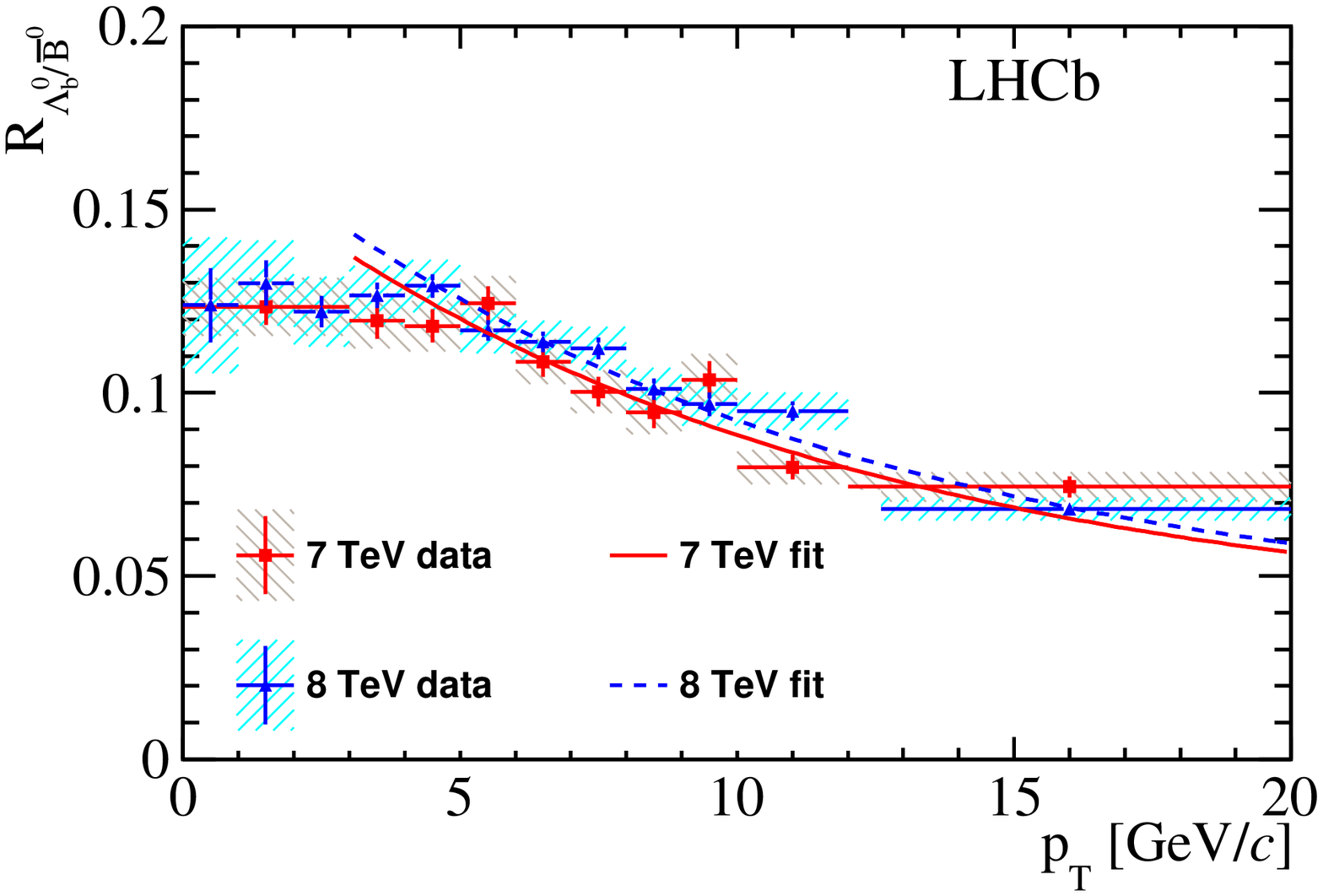}
 \includegraphics[width=0.49\linewidth]{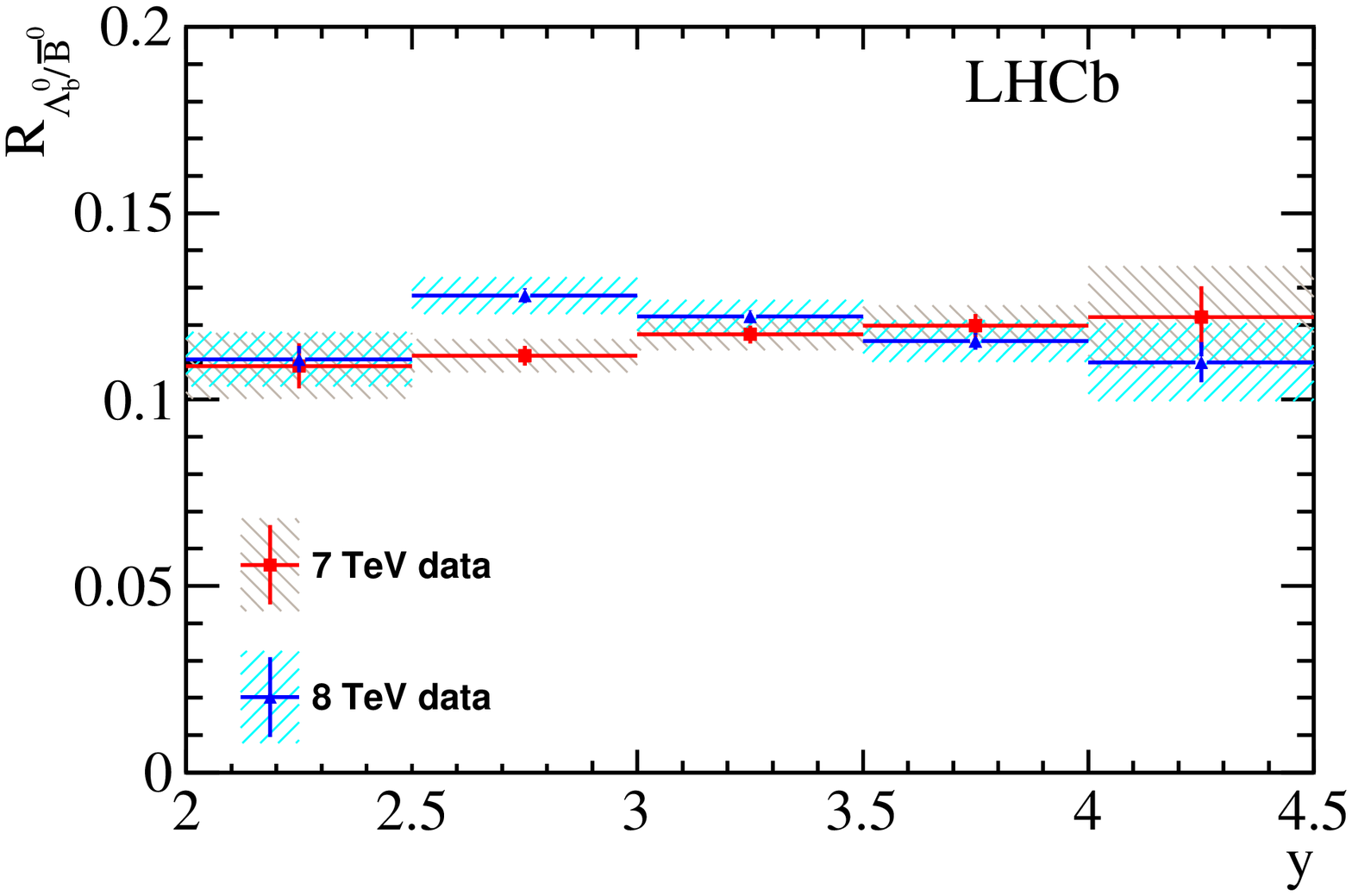}
\end{center}
\vspace*{-0.5cm}
\caption{
 Ratio $\RLbBdb$ as a function of (left) $\pt$ and (right) $y$ for the 2011 and 2012 
 samples, where the error bars indicate statistical uncertainties and 
 the hatched areas the total uncertainties. 
 The red solid (blue dashed) line in the left plot 
 represents the fit to the ratio $\fLbd(\pt)$ from Ref.~\cite{LHCb-PAPER-2014-004} 
 for the 2011 (2012) data sample.
 }\label{fig:ratio}
\end{figure}

The measured values of the ratio $\RLbBdb$, defined in Eq.~\ref{eq1}, 
as a function of \pt and \y are shown in Fig.~\ref{fig:ratio}.
In the region $\pt<5\gevc$, no \pt dependence of the ratio $\RLbBdb$ is observed, 
while the ratio decreases for $\pt>5\gevc$.
No dependence with rapidity is observed.
In Fig.~\ref{fig:ratio} the \pt dependence of the ratio \RLbBdb 
is fitted with the fragmentation function ratio $\fLbd(\pt)$ given in 
Ref.~\cite{LHCb-PAPER-2014-004}, which is only defined in the range $\pt>3\gevc$.

The asymmetry \apd between $\Lb$ and $\Lbbar$ is shown in Fig.~\ref{fig:asyresult}. The values are listed in Table~\ref{tab:asyresult} in the Appendix.
The data points are fitted with linear functions. 
The slope fitted to the asymmetry as a function of \pt is consistent with zero,
$(2.3\pm3.0)\times10^{-3}/(\mygevc)$ for $7\tev$ 
and $(3.5\pm2.0)\times10^{-3}/(\mygevc)$ for $8\tev$.
The fit to $\apd(y)$ gives a non-zero slope, 
and a combination of the results for $7\tev$ and $8\tev$ gives
\begin{equation*}
 \apd(\y)=(-0.001 \pm 0.007) + (0.058 \pm 0.014) (\y-\mean{y}),
\end{equation*}
where $\mean{y}=3.1$ is the average rapidity of $\Lb$ hadrons in the data sample. 
The non-zero slope suggests some baryon number transport from 
the beam particles to the less centrally produced \Lb, 
which leads to a $\Lb/\Lbbar$ cross-section ratio that increases with rapidity and which can be interpreted as, 
for example, 
a string drag effect or leading quark effect~\cite{Rosner:2014gta,Rosner:2012pi}. 

\begin{figure}[tb]
\begin{center}
 \includegraphics[width=0.49\linewidth]{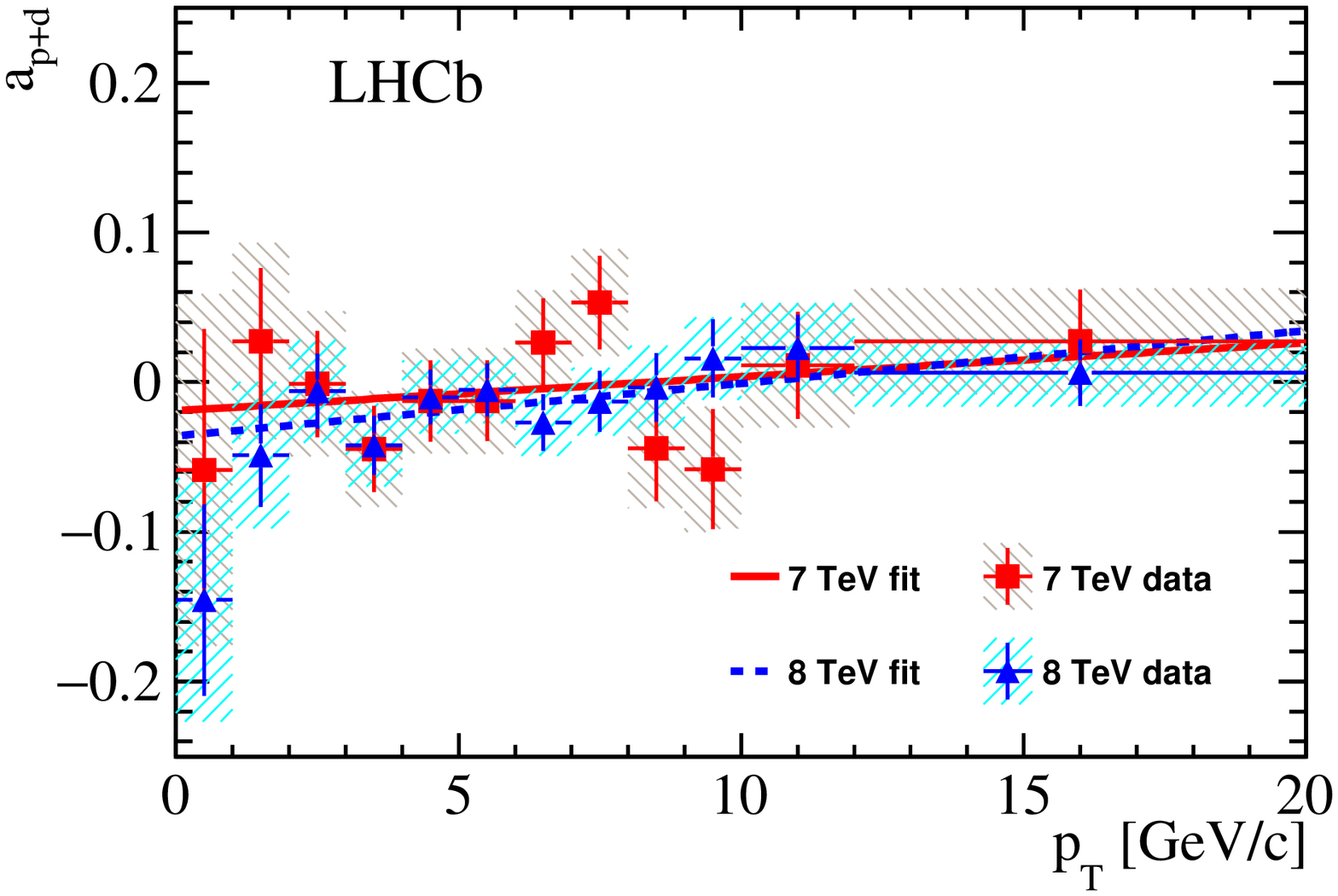}
 \includegraphics[width=0.49\linewidth]{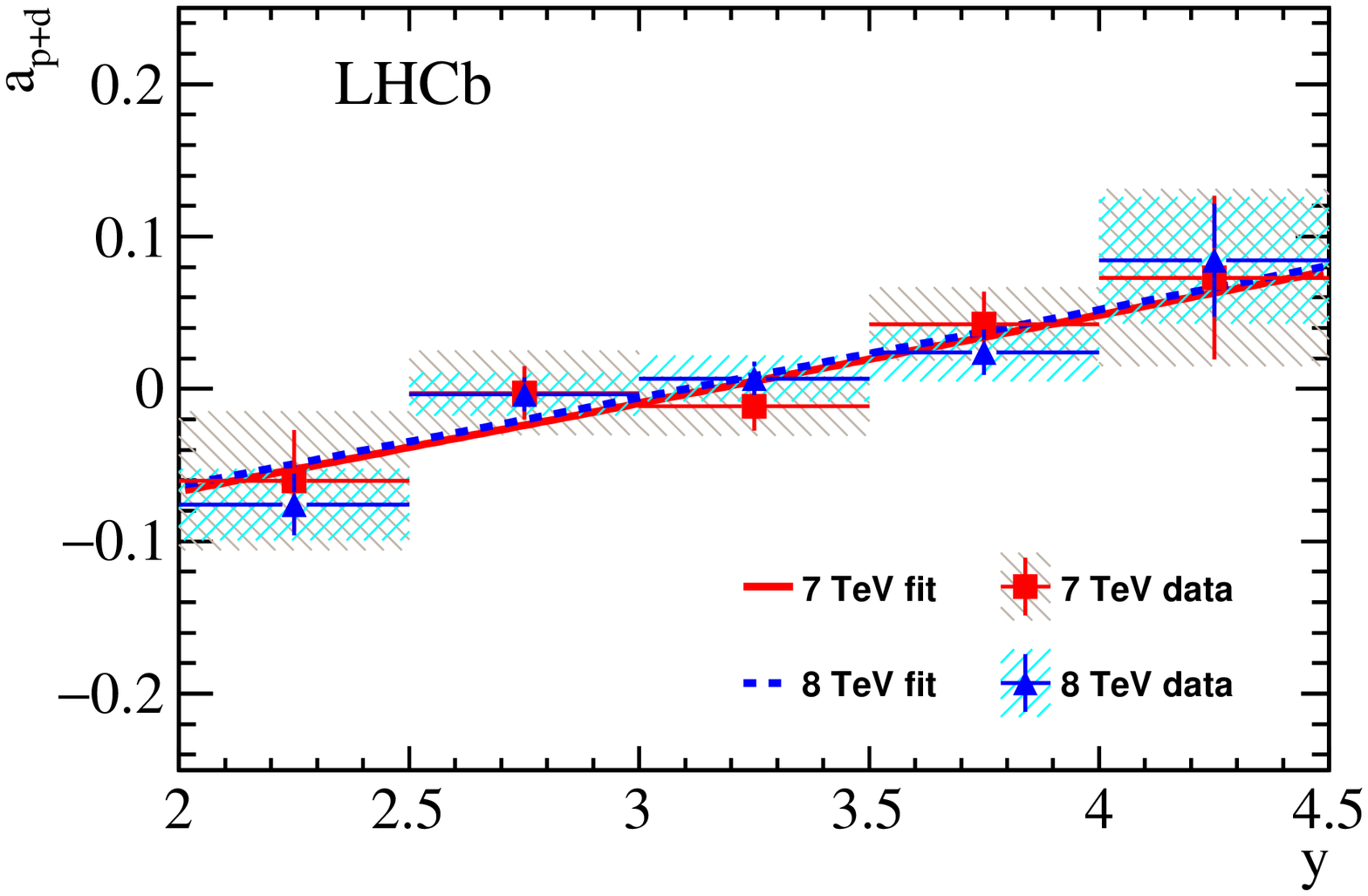}
\end{center}
\vspace*{-0.8cm}
\caption{
 Asymmetries \apd between $\Lb$ and $\Lbbar$ 
 as functions of (left) $\pt$ and (right) $y$. 
 The error bars indicate statistical uncertainties, and the hatched areas the total uncertainties.
 }
\label{fig:asyresult}
\end{figure}

\section{Branching fraction results}
The ratio $\RLbBdb$ can be calculated in bins of \pt as:
\begin{equation}
 \RLbBdb(\pt)=
 \frac{\NsigLb(\pt)~\epsBdb(\pt)}{\NsigBdb(\pt)~\epsLb(\pt)}~\BR(\KstarzbKpi).
\end{equation}
It is related to the fragmentation fraction ratio $\fLbd$ through
\begin{equation}
\label{eq4}
 \RLbBdb(\pt)=\frac{\BR(\Lbpk)}{\BR(\Bpik)}\fLbd(\pt)
 \equiv{}\mathcal{S}~\fLbd(\pt),
\end{equation}
where $\mathcal{S}\equiv\BR(\Lbpk)/\BR(\Bpik)$ is a constant factor, 
which can be determined from the fit in Fig.~\ref{fig:ratio}.
The absolute branching fraction of the decay $\Lbpk$ can then be measured as
\begin{equation}
\label{eq5}
 \BR(\Lbpk)=\mathcal{S}~\BR(\Bpik).
\end{equation}
The average of the fit results for the $7$ and $8\tev$ samples gives 
${\cal S} = 0.2458 \pm 0.0030$, 
which results in
\begin{equation*}
\BR(\Lbpk)=\input{table/number/branching_fraction/res}.
\end{equation*}
The first uncertainty is statistical, the second is systematic, 
the third is due to the uncertainty on the branching fraction of the $\Bpik$ decay,
and the fourth is due to the knowledge of $\fLbd$.

In Ref.~\cite{LHCb-PAPER-2014-020} the ratio $\BR(\LbJpsippi)/\BR(\Lbpk)$ was reported. 
Combining this with the value of $\BR(\Lbpk)$ above, the branching fraction 
of \LbJpsippi is determined as 
\begin{equation*}
 \BR(\LbJpsippi)=(2.61 \pm 0.09 \pm 0.13^{+0.47}_{-0.37})\times10^{-5},
\end{equation*}
where the first uncertainty is statistical, the second is due to the systematic uncertainty
on $\BR(\LbJpsippi)/\BR(\Lbpk)$,
and the third is due to systematic uncertainty on $\BR(\Lbpk)$.

Two pentaquark-charmonium states, $P_c(4380)^+$ and $P_c(4450)^+$, were observed by \lhcb
in the amplitude analysis of the $\Lbpk$ decay~\cite{LHCb-PAPER-2015-029},
and the fractions $f(P_c^+)$ of the two pentaquark-charmonium states 
in the $\Lbpk$ decay were measured.
Using these fractions and the value of $\BR(\Lbpk)$ obtained in this analysis,
the branching fractions $\BR(\Lb\to P_c^+\Km)\BR(P_c^+\to\jpsi\proton)$ are calculated as
\begin{eqnarray*}
\label{eqn:BR_Pc}
\BR(\Lb\to P_c^+\Km)\BR(P_c^+\to\jpsi\proton)
&=&f(P_c^+)\BR(\Lbpk)\\
&=&\begin{cases}
&(2.66\pm0.22\pm1.33^{+0.48}_{-0.38})\times10^{-5} \quad {\rm for}\quad P_c(4380)^+,\\
&(1.30\pm0.16\pm0.35^{+0.23}_{-0.18})\times10^{-5} \quad {\rm for}\quad P_c(4450)^+,
\end{cases}
\end{eqnarray*}
where the first uncertainty is statistical, the second is due to the systematic uncertainty
on $f(P_c^+)$, 
and the third is due to the systematic uncertainty on $\BR(\Lbpk)$.

\section{Conclusion}

Using a data sample corresponding to an integrated luminosity of  $3\invfb$ collected by the LHCb detector in 2011 and 2012,
the product of the \Lb differential production cross-section 
and the branching fraction of the decay $\Lbpk$ is measured
as a function of the $\Lb$ baryon's transverse momentum and rapidity.
The product of the \Bdb differential production cross-section
and the branching fraction of the decay $\Bpik$ is also measured.
The kinematic region of the measurements is $\pt<20\gevc$ and $2.0<y<4.5$.

The ratios of the cross-sections at $\sqs=8\tev$ to those at $\sqs=7\tev$ 
are calculated for \Lb and \Bdb hadrons and are compared with FONLL predictions.
The \pt dependence of the ratios is consistent with the FONLL calculations,
while the \y dependence is not consistent.
The production ratios of the $\Lb$ and $\Bdb$ hadrons are given 
for the 2011 and 2012 samples separately, and are consistent with
the dependence on $\pt$ and $y$ of the $b$ hadron observed in a previous LHCb analysis.
The asymmetry $\apd$ between $\Lb$ and $\Lbbar$ is also measured as a function of $\pt$ and $y$.
The result suggests some baryon number transport from the beam particles to the \Lb baryons.

Using information on the fragmentation ratio $\fLbd$ from a previous LHCb measurement,
the absolute branching fraction $\BR(\Lbpk)$ is obtained.
Using previous \lhcb measurements, the branching fractions
$\BR(\LbJpsippi)$ and $\BR(\Lb\to P_c^+\Km)\BR(P_c^+\to\jpsi\proton)$ are determined.

%% file: table/number/integral/Lb11/ans.tex
6.12\pm0.10\stat\pm0.25\syst\nb

%% file: table/number/integral/Lb12/ans.tex
7.51\pm0.08\stat\pm0.31\syst\nb

%% file: table/number/integral/B011/ans.tex
53.4\pm0.3\xx\stat\pm2.0\xx\syst\nb

%% file: table/number/integral/B012/ans.tex
63.6\pm0.2\xx\stat\pm2.3\xx\syst\nb

%% file: acknowledgements.tex
\section*{Acknowledgements}
\noindent We thank J. L. Rosner for interesting discussions of asymmetry in beauty baryon production.
We express our gratitude to our colleagues in the CERN
accelerator departments for the excellent performance of the LHC. We
thank the technical and administrative staff at the LHCb
institutes. We acknowledge support from CERN and from the national
agencies: CAPES, CNPq, FAPERJ and FINEP (Brazil); NSFC (China);
CNRS/IN2P3 (France); BMBF, DFG, HGF and MPG (Germany); INFN (Italy); 
FOM and NWO (The Netherlands); MNiSW and NCN (Poland); MEN/IFA (Romania); 
MinES and FANO (Russia); MinECo (Spain); SNSF and SER (Switzerland); 
NASU (Ukraine); STFC (United Kingdom); NSF (USA).
The Tier1 computing centres are supported by IN2P3 (France), KIT and BMBF 
(Germany), INFN (Italy), NWO and SURF (The Netherlands), PIC (Spain), GridPP 
(United Kingdom).
We are indebted to the communities behind the multiple open 
source software packages on which we depend. We are also thankful for the 
computing resources and the access to software R\&D tools provided by Yandex LLC (Russia).
Individual groups or members have received support from 
EPLANET, Marie Sk\l{}odowska-Curie Actions and ERC (European Union), 
Conseil g\'{e}n\'{e}ral de Haute-Savoie, Labex ENIGMASS and OCEVU, 
R\'{e}gion Auvergne (France), RFBR (Russia), XuntaGal and GENCAT (Spain), Royal Society and Royal
Commission for the Exhibition of 1851 (United Kingdom).

%% file: app_result.tex
\clearpage

\appendix
\vspace{-1.5cm}
{\noindent\bf\Large Appendix}

\begin{table}[H]
\caption{
 Products of \Lb production cross-sections (\mypb) and the branching fraction $\BR(\Lbpk)$
 in bins of $\pt$ and $y$ in the 2011 data sample.
The first uncertainties are statistical and the second are systematic.
}
\label{tab:final_result_Lb11}
\vspace*{-0.5cm}
\begin{center}
 \scalebox{1.0}{
\begin{tabular}{cccc}
\hline
\input{table/X/Lb11/production}
\hline
\end{tabular}
 }
\end{center}
\end{table}
\begin{table}[H]
\caption{
 Products of \Lb production cross-sections (\mypb) and the branching fraction $\BR(\Lbpk)$ 
 in bins of \pt and \y in the 2012 data sample.
 The first uncertainties are statistical and the second are systematic.
}
\label{tab:final_result_Lb12}
\vspace*{-0.5cm}
\begin{center}
 \scalebox{1.0}{
\begin{tabular}{cccc}
\hline
\input{table/X/Lb12/production}
\hline
\end{tabular}
 }
\end{center}
\end{table}

\begin{table}[H]
\caption{
    Products of $\Bdb$ production cross-sections (\mypb) and $\BR(\Bpik)$ 
    in bins of $\pt$ and $y$ in the 2011 data sample.
    The first uncertainties are statistical and the second are systematic.
}
\label{tab:final_result_B011}
\vspace*{-0.5cm}
\begin{center}
\begin{tabular}{cccc}
\hline
\input{table/X/B011/production}
\hline
\end{tabular}
\end{center}
\end{table}

\begin{table}[H]
\caption{
    Products of $\Bdb$ production cross-sections ($\mypb$) and $\BR(\Bpik)$ 
    in bins of $\pt$ and $y$ in the 2012 data sample.
    The first uncertainties are statistical and the second are systematic.
}
\label{tab:final_result_B012}
\vspace*{-0.5cm}
\begin{center}
\begin{tabular}{cccc}
\hline
\input{table/X/B012/production}
\hline
\end{tabular}
\end{center}
\end{table}

\begin{table}[H]
\caption{
    Asymmetries \apd $(\%)$ of $\Lb$ and $\Lbbar$ 
    in bins of $\pt$ and $y$ for the 2011 and 2012 samples.
    The first uncertainties are statistical and the second are systematic.
}
\label{tab:asyresult}
\vspace*{-0.5cm}
\begin{center}
\begin{tabular}{ccc}
\hline
\input{table/Asymmetry/ans_pt}
\hline
\input{table/Asymmetry/ans_y}
\hline
\end{tabular}
\end{center}
\end{table}

\clearpage

%% file: table/X/Lb11/production.tex
$\pt[\gevc]$	& $2.0<y<2.5$	& $2.5<y<3.0$	& $3.0<y<3.5$	\\
\hline
$0-3$ & $326\pm42\pm44$ & $354\pm18\pm23$ & $319\pm14\pm20$ \\ 
$3-4$ & $439\pm58\pm54$ & $503\pm27\pm33$ & $486\pm22\pm31$ \\ 
$4-5$ & $445\pm48\pm48$ & $425\pm21\pm27$ & $376\pm17\pm22$ \\ 
$5-6$ & $411\pm39\pm45$ & $297\pm15\pm17$ & $296\pm13\pm17$ \\ 
$6-7$ & $224\pm23\pm24$ & $235\pm12\pm14$ & $203\pm10\pm12$ \\ 
$7-8$ & $162\pm17\pm16$ & $175\pm\xx9\pm11$ & $145\pm7.4\pm9.2$	\\ 
$8-9$ & $100\pm12\pm\xx9$ & $109\pm6.5\pm7.0$ & $92.7\pm5.5\pm6.3$ \\ 
$9-10$   & $83.2\pm9.7\pm8.1$ & $93.6\pm6.0\pm6.4$ & $63.6\pm4.4\pm4.6$ \\ 
$10-12$     & $53.6\pm4.6\pm4.3$ & $39.5\pm2.3\pm2.4$ & $29.0\pm1.8\pm1.9$ \\ 
$12-20$     & $11.4\pm0.8\pm0.7$ & $11.3\pm0.6\pm0.6$ & $8.6\pm0.6\pm0.6$\\ 
\hline
              & $3.5<y<4.0$           & $4.0<y<4.5$ &	\\
\hline
$0-3$ & $244\pm13\pm 19$ 	 & $221 	\pm 26 	\pm 35$  &\\ 
$3-4$ & $371\pm21\pm 32$ 	 & $231 	\pm 29 	\pm 38$  &\\ 
$4-5$ & $294\pm16\pm 22$ 	 & $138 	\pm 18 	\pm 19$  &\\ 
$5-6$ & $229\pm12\pm 17$ 	 & $\xx95 	\pm 14 	\pm 16$  &\\ 
$6-7$ & $151\pm\xx9\pm12$ 	 & $\xx61 	\pm 11 	\pm \xx8$ &\\ 
$7-8$ & $99.0\pm6.5\pm8.1$ 	 & $38.4 	\pm 0.8 \pm 6.2$  &\\ 
$8-9$ & $69.0\pm5.3\pm 5.9$ 	 & $37.7 	\pm 7.7 \pm 5.9$  &\\ 
$9-10$   & $43.3\pm4.1\pm 4.2$ 	 & $22.8 	\pm 5.8 \pm 4.0$  &\\ 
$10-12$     & $20.4\pm1.9\pm 1.7$ 	 & $\xx7.8 	\pm 1.8 \pm 1.2$  &\\ 
$12-20$     & $\xx4.0\pm0.4\pm0.4$ 	 & $\xx2.2 	\pm 0.6 \pm 0.5$  &\\ 

%% file: table/X/Lb12/production.tex
$\pt[\gevc]$	& $2.0<y<2.5$	& $2.5<y<3.0$	& $3.0<y<3.5$	\\
\hline
$0-1$ & $100\pm29\pm33$      & $159\pm20  \pm27$     & $157\pm15 \pm 22$ \\ 
$1-2$ & $465\pm64\pm88$      & $487\pm29  \pm50$     & $433\pm24\pm 46$\\ 
$2-3$ & $661\pm63\pm120$     & $648\pm30  \pm58$     & $541\pm22\pm 41$ \\ 
$3-4$ & $706\pm51\pm94$      & $715\pm25  \pm52$     & $559\pm18 \pm 38$\\ 
$4-5$ & $579\pm39\pm68$      & $624\pm20  \pm39$     & $417\pm12\pm 27$\\ 
$5-6$ & $463\pm28\pm47$      & $446\pm14  \pm28$     & $356\pm10\pm 23$	\\ 
$6-7$ & $318\pm20\pm29$      & $322\pm10  \pm20$     & $210\pm\xx7\pm 12$ \\ 
$7-8$ & $248\pm15\pm23$      & $236\pm\xx8\pm15$     & $159\pm\xx5\pm 10$\\ 
$8-9$ & $173\pm11\pm18$      & $140.7\pm5.4\pm9.2$   & $118.4\pm 4.4\pm 7.8$\\ 
$9-10$   & $130\pm\xx9\pm13$    & $\xx92.6\pm3.9\pm6.3$    & $\xx65.4\pm 2.9\pm 4.4$\\ 
$10-12$     & $81.3\pm4.5\pm7.0$   & $\xx57.1\pm2.1\pm3.4$    & $\xx38.1\pm 1.5\pm 2.4$\\ 
$12-20$     & $15.2\pm0.7\pm1.0$ & $\xx13.7\pm0.5\pm0.8$ & $\xx\xx9.5\pm0.4\pm0.6$ \\ 
\hline
& $3.5<y<4.0$	& $4.0<y<4.5$	 &	\\
\hline
$0-1$ & $141 	\pm 18\pm 33$ 	   & $108 \pm 29 \pm 51$ 	\\ 
$1-2$ & $269 	\pm 20\pm 41$ 	   & $222 \pm 36 \pm 52$ 	\\ 
$2-3$ & $427 	\pm 21\pm 48$ 	   & $234 \pm 28 \pm 43$ 	\\ 
$3-4$ & $393 	\pm 17\pm34$ 	   & $256 \pm 25 \pm 45$ 	\\ 
$4-5$ & $324 	\pm 12\pm 27$ 	   & $195 \pm 17 \pm 26$ 	\\ 
$5-6$ & $229 	\pm \xx9\pm 16$	   & $111 \pm 11 \pm 16$ 	\\ 
$6-7$ & $152 	\pm \xx7\pm 11$   & $\xx99\pm 10 \pm 14$ 	\\ 
$7-8$ & $114   	\pm \xx5\pm\xx9$    & $51.3\pm5.8\pm 6.4$ 	\\ 
$8-9$ & $\xx74.7\pm 4.2 \pm 6.1$   & $30.8\pm 5.0 \pm 5.0$ 	\\ 
$9-10$   & $\xx55.4\pm 3.5 \pm 5.4$   & $17.4\pm 3.5 \pm 2.9$ 	\\ 
$10-12$     & $\xx27.7\pm 1.7 \pm 2.3$   & $10.3\pm 1.6 \pm 1.4$ 	\\ 
$12-20$     & $\xx\xx6.1 \pm 0.4 \pm 0.5$   & $\xx1.4\pm 0.4\pm 0.2$ 	\\ 

%% file: table/X/B011/production.tex
$\pt[\gevc]$	& $2.0<y<2.5$	& $2.5<y<3.0$	& $3.0<y<3.5$	\\
\hline
$0-3$ & $2850\pm130\pm200$     & $2870\pm50\pm140$     & $2580\pm30\pm110$\\ 
$3-4$ & $4420\pm220\pm340$     & $4350\pm80\pm220$     & $3740\pm60\pm170$\\ 
$4-5$ & $3540\pm160\pm250$     & $3570\pm60\pm160$     & $3310\pm50\pm150$ \\ 
$5-6$ & $2620\pm100\pm170$     & $2920\pm50\pm130$     & $2330\pm40\pm100$ \\ 
$6-7$ & $2290\pm\xx80\pm150$   & $2150\pm40\pm100$     & $1820\pm30\pm\xx80$\\ 
$7-8$ & $1790\pm\xx70\pm110$   & $1630\pm30\pm\xx80$   & $1320\pm20\pm\xx60$ \\ 
$8-9$ & $1260\pm\xx50\pm\xx80$ & $1150\pm20\pm\xx60$   & $\xx877\pm17\pm\xx42$ \\ 
$9-10$ & $\xx853\pm\xx34\pm\xx53$ & $\xx862\pm19\pm\xx43$ & $\xx613\pm14\pm\xx31$ \\ 
$10-12$ & $\xx581\pm\xx17\pm\xx32$   & $\xx540\pm10\pm\xx25$ & $\xx411\pm\xx8\pm\xx20$\\ 
$12-20$ & $\xx172\pm\xx\xx4\pm\xx\xx8$ & $\xx141\pm\xx2\pm\xx\xx6$ & $\xx102\pm\xx2\pm\xx\xx5$\\ 
\hline
	& $3.5<y<4.0$	& $4.0<y<4.5$	 &	\\
\hline
   $0-3$ & $2110\pm 30\pm\xx90$	 & $1450\pm 40 \pm \xx80$  &\\ 
   $3-4$ & $2660\pm 50\pm 130$ 	 & $1790\pm 70\pm 130$  &\\ 
   $4-5$ & $2310\pm 40\pm 110$ 	 & $1460\pm 60\pm 110$  &\\ 
   $5-6$ & $1750\pm 30\pm\xx80$  & $1050\pm 40 \pm \xx80$  &\\ 
   $6-7$ & $1190\pm 30\pm\xx60$  & $\xx608 \pm 30 \pm \xx48$  &\\ 
   $7-8$ & $\xx853\pm20\pm\xx45$  & $\xx573 \pm 29 \pm \xx51$  &\\ 
   $8-9$ & $\xx650\pm18\pm\xx37$  & $\xx385 \pm 21 \pm \xx38$  &\\ 
   $9-10$	 & $\xx424\pm14\pm\xx27$  & $\xx207 \pm 15 \pm \xx23$  &\\ 
   $10-12$	 & $\xx258\pm\xx7\pm\xx15$ & $\xx96\pm\xx6\pm\xx9$ &\\ 
   $12-20$	 & $\xx\xx64\pm\xx2\pm\xx\xx4$  & $\xx\xx26\pm\xx2\pm\xx\xx3$ &\\ 

%% file: table/X/B012/production.tex
$\pt[\gevc]$	& $2.0<y<2.5$	& $2.5<y<3.0$	& $3.0<y<3.5$	\\
\hline
$0-1$ & $1260\pm 110\pm 200$   & $1240\pm40\pm\xx80$ & $1100\pm 30 \pm\xx60$\\ 
$1-2$ & $3340\pm 170\pm 340$   & $3360\pm60\pm170$   & $3200\pm 50 \pm 150$ \\ 
$2-3$ & $5860\pm 220\pm 600$   & $4930\pm70\pm240$   & $4100\pm 50 \pm 180$ \\ 
$3-4$ & $6650\pm 200\pm 550$   & $5010\pm60\pm240$   & $4150\pm 50 \pm 180$ \\ 
$4-5$ & $4560\pm 120\pm 310$   & $4340\pm50\pm190$   & $3400\pm 40 \pm 140$ \\ 
$5-6$ & $4260\pm 100\pm 280$   & $3550\pm40\pm160$   & $2730\pm 30 \pm 120$  \\
$6-7$ & $2830\pm \xx60\pm 170$ & $2560\pm30\pm110$   & $1960\pm 20 \pm\xx80$   \\ 
$7-8$ & $2270\pm \xx50\pm 140$ & $1810\pm20\pm\xx80$    & $1460\pm20\pm\xx70$   \\
$8-9$ & $1650\pm \xx40\pm 100$ & $1460\pm20\pm\xx70$    & $1100\pm 10 \pm\xx50$   \\ 
$9-10$  & $1180\pm \xx30\pm\xx70$ & $1070\pm20\pm\xx50$    & $\xx696 \pm 10 \pm\xx34$   \\
$10-12$ & $\xx707\pm\xx13\pm\xx38$   & $\xx614\pm\xx7\pm\xx27$  & $489\pm\xx6\pm\xx23$   \\
$12-20$ & $\xx246\pm\xx\xx3\pm\xx11$ & $\xx180\pm\xx2\pm\xx\xx8$& $\xx129\pm\xx2\pm\xx\xx6$ \\
\hline
& $3.5<y<4.0$	& $4.0<y<4.5$	 &	\\
\hline
$0-1$ & $1010\pm30\pm\xx60$   & $\xx754\pm45\pm\xx73$ &\\
$1-2$ & $2830\pm50\pm150$     & $1720\pm 60\pm 140$  &\\
$2-3$ & $3200\pm50\pm160$     & $2460\pm 80\pm 200$  &\\
$3-4$ & $3180\pm40\pm150$     & $1770\pm 50\pm 130$  &\\
$4-5$ & $2610\pm30\pm120$     & $1650\pm 50\pm 120$  &\\
$5-6$ & $1900\pm30\pm\xx90$   & $1280\pm 40\pm 100$  &\\
$6-7$ & $1500\pm20\pm\xx70$   & $\xx816\pm26\pm\xx60$   &\\
$7-8$ & $1030\pm20\pm\xx50$   & $\xx621\pm22\pm\xx52$   &\\
$8-9$ & $\xx711\pm13\pm\xx38$ & $\xx390\pm15\pm\xx 34$  &\\
$9-10$ & $\xx478\pm10\pm\xx28$   & $\xx301\pm13\pm\xx 31$  &\\
$10-12$ & $\xx312\pm\xx6\pm\xx18$   & $\xx137\pm\xx5\pm\xx 12$  &\\
$12-20$ & $\xx\xx82\pm\xx2\pm\xx\xx5$ & $\xx\xx36\pm\xx2\pm\xx\xx3$  &\\

%% file: table/Asymmetry/ans_pt.tex
            & 2011                 & 2012 \\
\hline
\pt [\gevc] & & \\
\hline
$0-1$   & $-5.9\pm9.4\pm7.0$ & $-14.5\pm6.4\pm5.0$ \\
$1-2$   & $+2.7\pm4.9\pm4.4$ & $-4.9\pm3.4\pm3.5\xx$ \\
$2-3$   & $-0.1\pm3.6\pm3.3$ & $-0.6\pm2.5\pm2.4\xx$ \\
$3-4$   & $-4.5\pm2.9\pm2.6$ & $-4.2\pm2.0\pm1.9\xx$ \\
$4-5$   & $-1.3\pm2.7\pm2.2$ & $-1.1\pm1.8\pm1.7\xx$ \\ 
$5-6$   & $-1.3\pm2.7\pm2.3$ & $-0.5\pm1.8\pm1.3\xx$ \\
$6-7$   & $+2.7\pm2.9\pm1.9$ & $-2.7\pm1.9\pm1.2\xx$ \\
$7-8$   & $+5.3\pm3.1\pm1.7$ & $-1.3\pm2.0\pm1.0\xx$ \\
$8-9$   & $-4.4\pm3.5\pm1.9$ & $-0.3\pm2.3\pm1.6\xx$ \\
$9-10$  & $-5.8\pm4.0\pm1.2$ & $+1.6\pm2.6\pm0.9\xx$\\ 
$10-12$ & $+1.1\pm3.6\pm2.1$ & $+2.3\pm2.2\pm2.0\xx$ \\
$12-20$ & $+2.7\pm3.4\pm0.8$ & $+0.6\pm2.2\pm0.5\xx$ \\ 

%% file: table/Asymmetry/ans_y.tex
 $y$ & & \\
 \hline
$2.0<y<2.5$ & $-6.0\pm3.3\pm3.1$ & $-7.6\pm2.0\pm1.2$ \\
$2.5<y<3.0$ & $-0.3\pm1.7\pm2.1$ & $-0.3\pm1.1\pm0.9$ \\
$3.0<y<3.5$ & $-1.0\pm1.6\pm1.1$ & $+0.7\pm1.1\pm1.0$ \\
$3.5<y<4.0$ & $+4.3\pm2.1\pm1.1$ & $+2.4\pm1.5\pm1.2$ \\
$4.0<y<4.5$ & $+7.3\pm5.4\pm2.2$ & $+8.4\pm3.7\pm1.9$ \\ 

%% file: LHCb_HD_authorlist_2015-06-23.tex
\centerline{\large\bf LHCb collaboration}
\begin{flushleft}
\small
R.~Aaij$^{38}$, 
B.~Adeva$^{37}$, 
M.~Adinolfi$^{46}$, 
A.~Affolder$^{52}$, 
Z.~Ajaltouni$^{5}$, 
S.~Akar$^{6}$, 
J.~Albrecht$^{9}$, 
F.~Alessio$^{38}$, 
M.~Alexander$^{51}$, 
S.~Ali$^{41}$
% ...

{\footnotesize \it
% ...
$ ^{5}$Clermont Universit\'{e}, Universit\'{e} Blaise Pascal, CNRS/IN2P3, LPC, Clermont-Ferrand, France\\
$ ^{6}$CPPM, Aix-Marseille Universit\'{e}, CNRS/IN2P3, Marseille, France\\
% ...
$ ^{9}$Fakult\"{a}t Physik, Technische Universit\"{a}t Dortmund, Dortmund, Germany\\
% ...
$ ^{37}$Universidad de Santiago de Compostela, Santiago de Compostela, Spain\\
$ ^{38}$European Organization for Nuclear Research (CERN), Geneva, Switzerland\\
% ...
$ ^{41}$Nikhef National Institute for Subatomic Physics, Amsterdam, The Netherlands\\
% ...
$ ^{46}$H.H. Wills Physics Laboratory, University of Bristol, Bristol, United Kingdom\\
% ...
$ ^{51}$School of Physics and Astronomy, University of Glasgow, Glasgow, United Kingdom\\
$ ^{52}$Oliver Lodge Laboratory, University of Liverpool, Liverpool, United Kingdom\\
% ...
}
\end{flushleft}